%% file: Main_v5.2.tex
\documentclass[
  reprint,                % two-column PRL style
  superscriptaddress,
  amsmath,amssymb,
  aps,
  prl,                    % target journal
  floatfix
]{revtex4-2}
\usepackage{ragged2e} % adds \justifying
\usepackage[justification=justified,singlelinecheck=false]{caption}
\usepackage{graphicx}     % include figures
\usepackage{dcolumn}      % align table columns on decimal point
\usepackage{bm}           % bold math
\usepackage[colorlinks=true,
            citecolor=blue,
            linkcolor=magenta,
            urlcolor=blue]{hyperref}
\usepackage[utf8]{inputenc}
\usepackage[normalem]{ulem}
\usepackage{siunitx}
\usepackage[protrusion=true,expansion=true]{microtype}

\usepackage[colorlinks=true,citecolor=blue,linkcolor=magenta,urlcolor=blue]{hyperref}
\usepackage{xr-hyper}
\newcommand{\YBCO}{YBa$_2$Cu$_3$O$_{7-\delta}$} % Choose template
\begin{document}

%\title{Relieved Polar Catastraphe on YBa$_2$Cu$_3$O$_{7-\delta}$ surface with Pr doping}
\title{Superconductivity suppression and bilayer decoupling in Pr substituted \YBCO}

\author{Jinming Yang$^{*}$}
\affiliation{Department of Physics, Yale University, New Haven, CT 06511,USA}
\affiliation{Department of Applied Physics, Yale University, New Haven, CT 06511, USA}
\author{Zheting Jin$^{*}$}
\affiliation{Department of Applied Physics, Yale University, New Haven, CT 06511, USA}
\author{Siqi Wang$^{*}$}
\affiliation{Department of Applied Physics, Yale University, New Haven, CT 06511, USA}
\author{Camilla Moir$^{*}$}
\affiliation{Department of Physics, University of California, San Diego, CA 92093, USA}
\author{Mingyu Xu}
\affiliation{Department of Chemistry, Michigan State University, east lansing, MI 48824, USA}
\author{Brandon Gunn}
\affiliation{Department of Physics, University of California, San Diego, CA 92093, USA}
\author{Xian Du}
\affiliation{Department of Applied Physics, Yale University, New Haven, CT 06511, USA}
\author{Zhibo Kang}
\affiliation{Department of Applied Physics, Yale University, New Haven, CT 06511, USA}
\author{Keke Feng}
\affiliation{Department of Physics, University of California, San Diego, CA 92093, USA}
\author{Makoto Hashimoto}
\affiliation{Stanford Synchrotron Radiation Lightsource, Stanford Linear Accelerator Center (SLAC) National Accelerator Laboratory, Menlo Park, CA 94025, USA}
\author{Donghui Lu}
\affiliation{Stanford Synchrotron Radiation Lightsource, Stanford Linear Accelerator Center (SLAC) National Accelerator Laboratory, Menlo Park, CA 94025, USA}
\author{Jessica L McChesney}
\affiliation{Advanced Photon Source, Argonne National Laboratory, Lemont, IL 60439, USA}
\author{Martin Sundermann}
\affiliation{PETRA III, Deutsches Elektronen-Synchrotron DESY, Notkestraße 85, 22607 Hamburg, Germany}
\affiliation{Max Planck Institute for Chemical
Physics of Solids, Nöthnitzer Straße 40, 01187 Dresden, Germany}
\author{Hlynur Gretarsson}
\affiliation{PETRA III, Deutsches Elektronen-Synchrotron DESY, Notkestraße 85, 22607 Hamburg, Germany}
\author{Shize Yang}
\affiliation{Energy Sciences Institute, Yale University, West Haven, CT 06516, USA}
\author{Wei-Wei Xie}
\affiliation{Department of Chemistry, Michigan State University, east lansing, MI 48824, USA}
\author{Alex Frano}
\affiliation{Department of Physics, University of California, San Diego, CA 92093, USA}
\author{Sohrab Ismail-Beigi}
\affiliation{Department of Applied Physics, Yale University, New Haven, CT 06511, USA}
\author{M. Brian Maple}
\email{mbmaple@ucsd.edu}
\affiliation{Department of Physics, University of California, San Diego, CA 92093, USA}
\author{Yu He}
\email{yu.he@yale.edu}
\affiliation{Department of Applied Physics, Yale University, New Haven, CT 06511, USA}

\thanks{$^{*}$These authors contributed equally to this work.}
%\authorcontributions{Y.H., S.I-.B. designed research; J.Y., S.W., Z.J., Z.K., X.D. performed research; J.Y., Z.J. S.W. analyzed data; M.H., D. L., J.L.M. developed instruments for the experiment; C.M., A.F., M.X., B.G., W.W.X., M.B.M. synthesized and characterized samples; Z.J., S.I-.B. performed theoretical calculations; all authors wrote the paper.}
%\authordeclaration{Please declare any competing interests here.}
%\equalauthors{\textsuperscript{1}J.Y., Z.J. S.W., C.M. contributed equally to this work}
%\correspondingauthor{\textsuperscript{2}To whom correspondence should be addressed. E-mail: author.two\@email.com}

% At least three keywords are required at submission. Please provide three to five keywords, separated by the pipe symbol.
%\keywords{superconductivity $|$ cuprate $|$ photoemission $|$ rare earth element}

\begin{abstract}
The mechanism behind superconductivity suppression induced by Pr substitutions in \YBCO~(YBCO) has been a mystery since its discovery: in spite of being isovalent to Y$^{3+}$ with a small magnetic moment, it is the only rare-earth element that has a dramatic impact on YBCO's superconducting properties.
Using angle-resolved photoemission spectroscopy (ARPES) and DFT+$U$ calculations, we uncover how Pr substitution modifies the low-energy electronic structure of YBCO. 
Contrary to the prevailing Fehrenbacher–Rice (FR) and Liechtenstein–Mazin (LM) models, the low energy electronic structure contains no signature of any $f$-electron hybridization or new states. Yet, strong electron doping is observed primarily on the antibonding Fermi surface. Meanwhile, we reveal major  electronic structure modifications to Cu-derived states with increasing Pr substitution: a pronounced CuO$_2$ bilayer decoupling and an enhanced CuO chain hopping, implying indirect electron-release pathways beyond simple 4$f$ state ionization. Our results challenge the long-standing FR/LM mechanism, and establish Pr substituted YBCO as a potential platform for exploring correlation-driven phenomena in coupled 1D–2D systems.
\end{abstract}
%\dates{This manuscript was compiled on \today}
%\doi{\url{www.pnas.org/cgi/doi/10.1073/pnas.XXXXXXXXXX}}
\maketitle
%\thispagestyle{firststyle}
%\ifthenelse{\boolean{shortarticle}}{\ifthenelse{\boolean{singlecolumn}}{\abscontentformatted}{\abscontent}}{}
%\firstpage[14]{2}

Elemental substitution is a powerful route to tune superconductivity in cuprate superconductors. In the pair breaking theory of Abrikosov and Gor'kov, the depression of the superconducting transition temperature ($T_c$) with concentration of RE solute in a conventional spin-singlet superconductor is predicted to scale with the de-Gennes factor of the RE ion and the square of the strength of the exchange interaction between the localized moments and conduction electron spins~\cite{osti_4097498,maple1970dependence,maple1976superconductivity}. Surprisingly, the substitution of most rare-earth elements for Y in YBCO has little effect on $T_c$~\cite{YANG1987515} (Fig.~\ref{fig:Fig1_SCXRD}), suggesting that magnetic pair-breaking is very weak in these systems. Praseodymium (Pr) stands out as a striking exception: even partial substitution of Pr for Y leads to a rapid suppression of superconductivity~\cite{PhysRevLett.58.1891,liang1987,DALICHAOUCH19881001}. Pr is unique among the rare-earth series in having both one of the smallest de Gennes factors and a putatively less localized 4$f$ state, implying stronger hybridization with the conduction electrons~\cite{PhysRevB.40.5300,brown1987,PhysRevB.40.4453,MAPLE1992135}. This stronger hybridization motivated extensions of the original pair-breaking theory to include hybridization-induced exchange interactions, but even these models fail to fully explain the unusually strong suppression of $T_c$ by Pr substitution~\cite{MAPLE1992135,XU1992104,HShakeripour_2001,YAMANI199678}. Understanding why Pr substitution is uniquely destructive to superconductivity thus remains a key open question, and resolving this puzzle may provide new insights into the pairing mechanism in cuprates.

In view of the inability of magnetic pair breaking to fully account for the strong depression of $T_c$ of Pr substituted YBCO, the depression of $T_c$ has been attributed in large part to the depletion of hole carriers.  
Thus, Pr substituted YBCO appears to be an underdoped system similar to oxygen-deficient YBCO, a viewpoint that is supported by evidence for the formation of a pseudogap in transport measurements~\cite{MAPLE1992135,YAMANI199678,PhysRevB.38.2910,PhysRevB.40.4517,PhysRevB.47.6043,PhysRevB.55.R3390, 10.1063/1.2215374} and Ca$^{2+}$ counter substitution experiments~\cite{PhysRevLett.63.2516,MAPLE199264}. However, electron doping appears unlikely due to the nominal isovalence of Pr$^{3+}$ and Y$^{3+}$. Subsequently, hole localization was proposed~\cite{Neukirch_1988,alleno1999,rosen1988,PhysRevB.41.8955,PhysRevB.42.4823,PhysRevB.49.535,PhysRevB.44.2410,SODERHOLM1989121,Lei_1998}. Such localization of holes could occur when Pr 4$f_{z(x^2-y^2)}$ orbitals are hybridized with O 2p$_\pi$ states to form new hole bands at the Fermi level ($E_F$), as suggested by FR~\cite{PhysRevLett.70.3471} and LM~\cite{PhysRevLett.74.1000} models. The models were based on the hypothesized distinguishing feature of Pr among rare earth elements: its 4$f_{z(x^2-y^2)}$ states lie at the $E_F$; however, direct experimental evidence for such low-energy 4$f$ contributions remains lacking.

Recently, Pr substituted YBCO was found to host long-range 3-dimensional charge order (CO)~\cite{ruiz2022} with in-plane CO at the Mott limit~\cite{doi:10.1073/pnas.2302099120}, where FR and LM model pictures are proposed to be relevant. Moreover, rare-earth infinite-layer nickelates \textit{RE}NiO$_2$ -- sharing structural similarities with Pr substituted YBCO -- exhibit parallel debates about $f$-state involvement~\cite{Wang2023}, but first-principles calculations indicate the absence of low-energy $f$ states~\cite{Liao2023}. These developments underscore the critical need to resolve how Pr substitution modifies the electronic structure in these archetypal superconducting transition metal oxides. Here, we employ angle-resolved photoemission spectroscopy (ARPES) to directly probe the low-energy electronic structure, which is complemented by density functional theory (DFT)+$U$ calculations and non-resonance inelastic x-ray scattering. Together, these techniques allow us to address the following key questions: i) whether the potential roles of $f$-state actively contributes to transport at the Fermi level; ii) how Pr substitution alters the low-energy electronic structure; and iii) if $T_c$ is indeed dictated by hole localization in Pr substituted YBCO.

\section*{Results}
\begin{figure}%[t!]
\centering
\includegraphics[width=\linewidth]{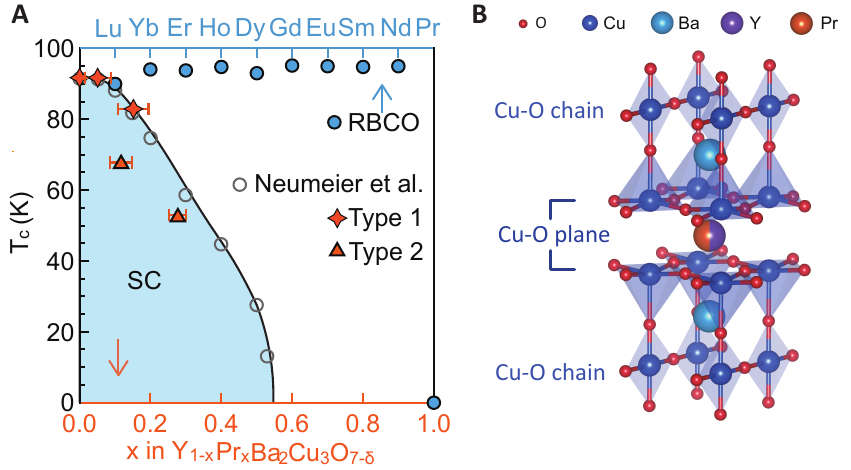}
\caption{\justifying Pr substituted YBCO and REBa$_2$Cu$_3$O$_{7-\delta}$ (RBCO) superconducting transition and crystal structure of Pr substituted YBCO. (A) Rare earth substitution effects in YBCO. Blue circles: RBCO superconducting transition temperature (adapted from ~\cite{YANG1987515}). Grey circles (adapted from ~\cite{NEUMEIER1992158}), orange stars (Type 1), and triangles (Type 2) : Pr substitution dependence of $T_c$. (B) Crystal structure of Pr substituted YBCO.}
\label{fig:Fig1_SCXRD}

\end{figure}

We investigate pristine YBCO single crystals ($T_c$ = 91 K) alongside four Pr substituted variants with $T_c$'s of 91 K, 84 K, 63 K, and 53 K, where Pr content is found to be 5 \%, 15 \%, 12\%, and 28\% by energy-dispersive X-ray spectroscopy (EDX) measurements, respectively. As shown in Fig.~\ref{fig:Fig1_SCXRD}A, samples measured are cataloged into two types. Type 1 follows the typical Pr substituted YBCO phase diagram, while type 2 shows lower $T_c$ at the same Pr content. The existence of type 2 implies potential additional $T_c$ suppression mechanisms in these samples. To elucidate the superconductivity suppression mechanism, we conduct ARPES measurements of the electronic structure.

\begin{figure*}%[B]
\centering
\includegraphics[width=17.4cm]{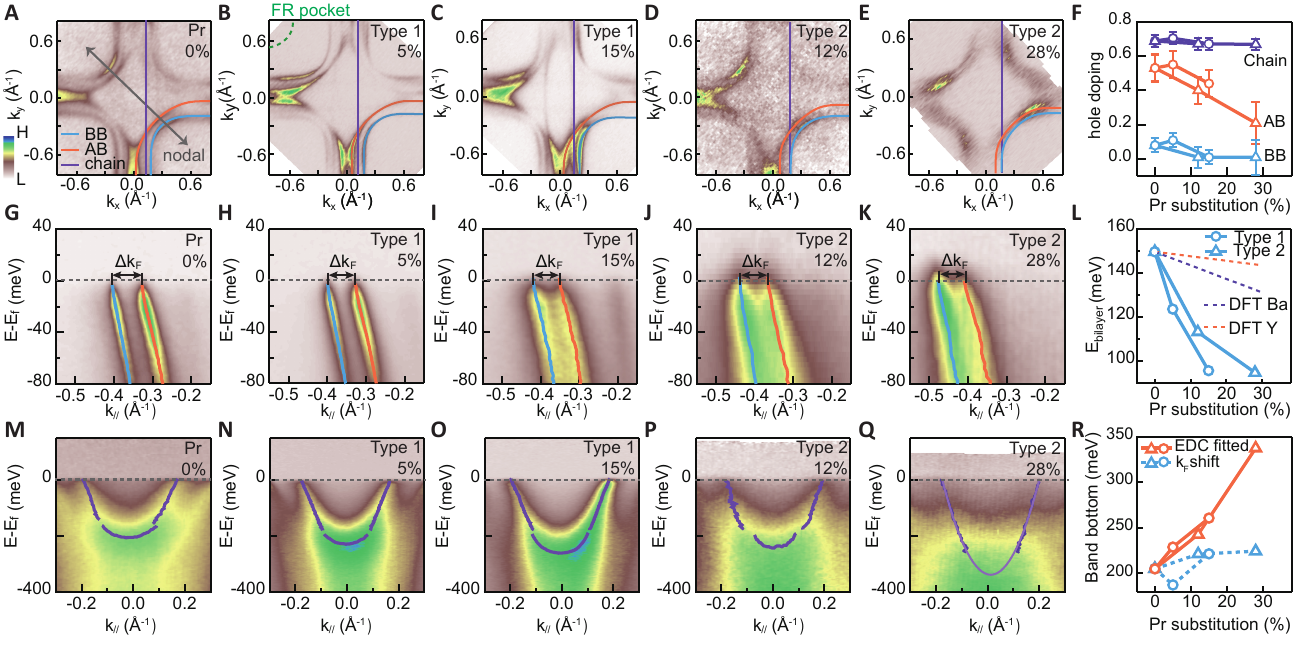}
\captionsetup{width=\linewidth}
\caption{\justifying ARPES measured electronic structure of Pr substituted YBCO. (A-E) Fermi surfaces with extracted Fermi momentum . (F) Hole doping level evolution for AB, BB and chain with Pr substitution. Circles (triangles) represent type 1 (type 2) samples. (G-K) Nodal cuts at 51 eV highlighting plane bands with fitted dispersions. (L) Bilayer splitting energy evolution with Pr substitution. (M-Q) Nodal cuts at 43 eV highlighting chain band with fitted dispersions. (R) Band bottom position evolution with Pr substitution. Circles (triangles) represent type 1 (type 2) samples. Orange curves are extracted directly from EDC fitting. Blue dashed lines are band bottom position evolution expected from pure charge doping effects.}
\label{fig:Fig2_ARPES}
\end{figure*}

Pristine YBCO exhibits three bands crossing $E_F$: the bonding (BB) and antibonding (AB) bands from the CuO$_2$ bilayer, and a quasi-1D band from the CuO chain~\cite{campuzano1991,PhysRevB.44.2399,mante1991,PhysRevB.45.5614,schabel1998a,PhysRevB.57.6090,PhysRevLett.86.4370,borisenko2006,PhysRevB.76.064519,dahm2009,PhysRevB.79.140503,hossain2008,PhysRevB.79.144528,fournier2010,zabolotnyy2012,Iwasawa2018,XingjiangYBCO}. By exploiting the photoemission dipole transition matrix element effects (see Fig.~\ref{supp-fig:Kz_nodal}), we selectively enhance either plane-derived or chain-derived bands for detailed analysis. In FR/LM models, Pr $4f$-O $2p_{\sigma}$ hybridization should generate an additional dispersive band with a hole pocket centered at ($\pi,\pi$) (dashed line, Fig.~\ref{fig:Fig2_ARPES}B)~\cite{PhysRevLett.74.1000}. However, ARPES measurements (Fig.~\ref{fig:Fig2_ARPES}B-E) show no such Pr-derived pocket at the Brillouin zone corner for all samples with up to 28\% Pr substitution. This absence is further confirmed by nodal cuts, which reveal no low-energy $4f$-related bands. Neither is there any visible indication of $f$-hybridization-induced anti-crossing near the Fermi level $E_F$ ( Fig.~\ref{supp-fig:Bilayer_nodal}). 

Hole reduction to the CuO$_2$ plane within the FR/LM picture occurs via hole transfer to the local Pr-O bonds. The absence of low-energy $4f$ bands implies that any hole doping reduction must directly occur on CuO$_2$ planes or CuO chains. Hole concentration changes are quantified by fitting a global tight-binding model to both BB and AB bands in the 0–50 meV binding energy range~\cite{chen2022c} (also see Fig.~\ref{supp-fig:TBMfitting} and Text S1). The model is defined as:
\begin{align*}
E_\pm &= \epsilon_{\pm} - 2t_\pm (\cos{k_x} + \cos{k_y}) \\
    &- 4t_\pm^\prime \cos{k_x} \cos{k_y} - 4t_\pm^{\prime\prime} (\cos{2k_x} + \cos{2k_y}) 
    \label{eqn:TBmodel}
\end{align*}
where $\pm$ denotes the bonding (BB) ($-$) or antibonding (AB) ($+$) band (see Text S1 for details).
Hole doping levels for the CuO$_2$ planes and CuO chains are extracted by applying Luttinger's theorem to their respective Fermi surface sheets. Previous ARPES studies of pristine YBCO were confounded by surface hole-doping saturation effects~\cite{PhysRevB.44.2399,mante1991,schabel1998a,PhysRevB.57.6090,PhysRevLett.86.4370,PhysRevB.76.064519,PhysRevB.79.140503,hossain2008,PhysRevB.79.144528,fournier2010,Iwasawa2018,XingjiangYBCO}, with electron doping only observed through surface alkali metal dosing~\cite{hossain2008,fournier2010}. Remarkably, in as-cleaved Pr substituted YBCO, we observe major electron doping effects manifested through a systematic shrinkage of hole pockets from the AB/BB bands (Fig.~\ref{fig:Fig2_ARPES}A-E). Fig.~\ref{fig:Fig2_ARPES}F reveals that electron doping occurs primarily on the AB band and moderately on the BB band of the CuO$_2$ bilayer, while the chain band shows minimal change. Notably, type 2 samples experience greater electron doping than type 1 samples, despite lower global Pr concentration, which unexpectedly leads to exacerbated $T_c$ suppression.

The disproportionate hole-doping on AB/BB bands also implies an appreciable change to the CuO$_2$ bilayer energy splitting with Pr substitution (Fig.~\ref{fig:Fig2_ARPES}A-E). Along the nodal direction (Fig.~\ref{fig:Fig2_ARPES}G-K), bilayer splitting energy measured with $\hbar v_F \Delta k_F$ is substantially suppressed by over 30\% from pristine to both types of Pr substituted YBCO (Fig.~\ref{fig:Fig2_ARPES}L), signaling rapid electronic decoupling of the CuO$_2$ bilayer. While first-principles calculations (dashed lines, Fig.~\ref{fig:Fig2_ARPES}L) show moderate decoupling depending on different Pr substitution sites, the experimental reduction exceeds theoretical predictions even more, indicating possible additional decoupling mechanisms.

\begin{figure*}%[H]
\centering
\includegraphics[width=17.4cm]{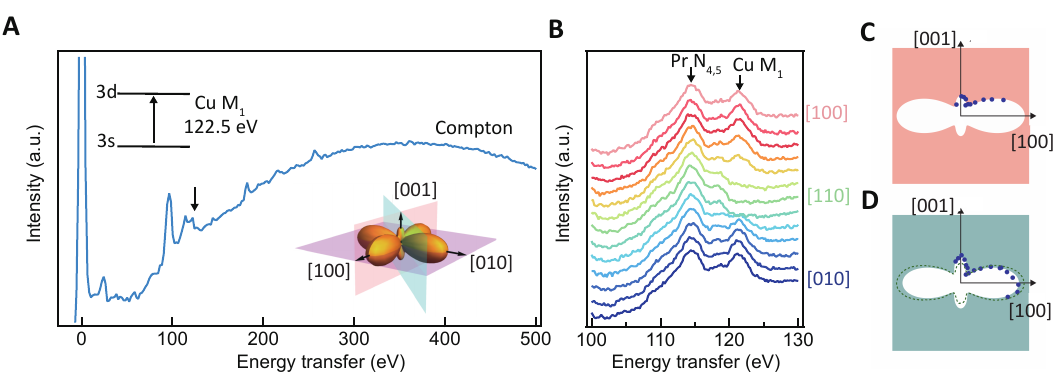}
\captionsetup{width=\linewidth}
\caption{\justifying Cu 3$d$ orbital imaging with non-resonance inelastic x-ray scattering. (A) Energy loss spectrum measured with an incident energy of 9690 eV on nominally 80\% Pr substituted \YBCO~ (non-superconducting) at 20~K along [100] direction. (B) In-plane angular dependence of the Cu M$_1$ transition intensity as a direct measure of the in-plane angular profile of Cu 3$d_{x^2-y^2}$ orbital. Orbital lobe profile along the $xz$ plane for (C) 80\% Pr substituted \YBCO~and (D) fully oxygenated pristine \YBCO ~($T_c$=90~K). Black dotted line denotes the orbital shape of Pr-doped sample for ease of comparison.}
\label{fig:Fig3_orbital}
\end{figure*}

Unlike the planar bands, the CuO-chain band shows only weak electron doping (purple lines, Fig.~\ref{fig:Fig2_ARPES}F). However, dramatic changes in the effective mass are observed. As shown in Fig.~\ref{fig:Fig2_ARPES}M-Q, chain bands steepen progressively with Pr substitution. This is characterized by a systematic shift of the chain band bottom toward higher binding energy. As shown in Fig.~\ref{fig:Fig2_ARPES}R, this energy shift (orange) far exceeds what the small electron doping can induce. Notably, quasiparticle coherence -- especially on the chain band -- is suppressed in type 2 samples (Fig.~\ref{fig:Fig2_ARPES}M-Q), suggesting disproportionately high Pr-induced disorder effects on the chain band, which will be discussed later.

To understand how the $e_g$ orbital of the planar copper evolves with Pr-substitution of Ca, we perform angle-dependent non-resonance inelastic x-ray scattering to trace the relative anisotropy of the in-plane (3$d_{x^2-y^2}$) and the out-of-plane (3$d_{z^2}$) orbitals. This technique is based on the principle that the angle-dependent transition probability for the dipole-forbidden 3$s$ to 3$d$ excitation (M$_1$ edge) exactly follows the angular profile of the 3$d$ orbital~\cite{Yavas2019}. As shown in Fig.~\ref{fig:Fig3_orbital}A, the Cu M$_1$ and Pr N$_{4,5}$ transitions are clearly resolved around 120 eV energy loss. Notably, the in-plane angular dependence of the Cu M$_1$ transition intensity follows the lobe-structure of the 3$d_{x^2-y^2}$ hole state (Fig.~\ref{fig:Fig3_orbital}B). By comparing the out-of-plane angular dependence of the M$_1$ transition probability in 80\% Pr substituted (Fig.~\ref{fig:Fig3_orbital}C) and pristine \YBCO~(Fig.~\ref{fig:Fig3_orbital}D), a clear `flattening' of overall Cu $e_g$ orbital occurs with heavy Pr-doping. This indicates lower energy and fewer hole states for the 3$d_{z^2}$ orbital with Pr substitution, henceforth offering a direct wavefunction view of the associated bilayer-decoupling, consistent with the bilayer splitting energy reduction seen in ARPES.

\section*{Discussion}
\begin{figure}%[H]
\centering
\includegraphics[width=\linewidth]{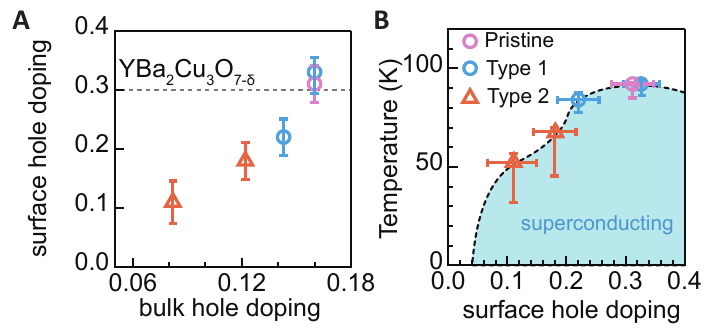}
\caption{\justifying Electron doping and superconductivity suppression in Pr substituted YBCO. (A) Surface hole doping level dependence on the bulk doping level for Pr substituted YBCO (markers) and YBCO with hole doping controlled by oxygen content (gray dashed line from ~\cite{PhysRevB.76.064519}). The bulk hole doping level for Pr substituted YBCO is obtained by comparing $T_c$ with the oxygen content controlled YBCO phase diagram in ~\cite{PhysRevB.73.180505}. (B) Superconducting transition temperature against surface hole doping level derived in this work.}
\label{fig:Fig4_results}
\end{figure}

Our observation suggests that electron doping to the CuO$_2$ plane is the major mechanism behind superconductivity suppression in the system. First, we note that the surface doping extracted by ARPES here is not directly comparable to bulk doping due to the lack of neutral cleavage planes in YBCO~\cite{PhysRevB.44.2399,PhysRevB.45.5614,PhysRevB.57.6090,PhysRevLett.86.4370,XingjiangYBCO,Iwasawa2018}\footnote{Charge-neutral cleavage planes have recently been observed, but only in small domain size on the order of several micrometer~\cite{XingjiangYBCO}. In this work, we focus on the typical cleavage plane domain that highlights the plane bands illustrated in ~\cite{Iwasawa2018}}. Such a polar surface will undergo charge redistribution to avoid the polar catastrophe~\cite{nakagawa2006}. As a result, the surface hole doping level of cleaved YBCO was found to be a constant at around 0.3 regardless of its oxygenation level~\cite{PhysRevB.79.064519}. 
The additional hole doping compared to the bulk is believed to be associated with the chain band charging state~\cite{PhysRevB.75.014513,PhysRevB.76.064519,PhysRevB.79.140503,PhysRevB.79.144528,XingjiangYBCO}. 
In remarkable contrast, on the as-cleaved surface for Pr substituted YBCO, the surface hole doping level changes accordingly with the bulk doping level (Fig.\ref{fig:Fig4_results}A).
%Given that the chain doping remains almost unchanged with Pr substitution, the surface hole doping extracted here can be qualitatively related to bulk doping by a proportionate offset. %
Fig.~\ref{fig:Fig4_results}B plots $T_c$ against the ARPES-derived surface hole doping, which qualitatively tracks the canonical bulk phase diagram. We emphasize that this replication of the dome shape from directly ARPES-derived hole doping levels strongly suggests that, regardless of the sample types, electron doping dominates the $T_c$ suppression. Intriguingly, the long-range c-axis charge order then occurs at a Pr substitution (30\%) qualitatively close to the equivalent charge order hole doping in oxygenated YBCO.
\begin{figure}[t]
\centering
\includegraphics[width=\linewidth]{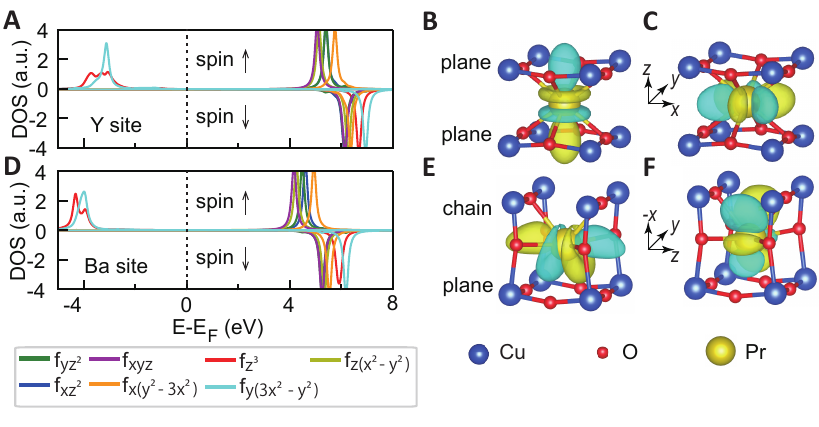}
\caption{\justifying Pr $f$-orbitals predicted by DFT calculations. (A) The ground-state projected density of state (DOS) of Pr $f$-orbitals on the Y site.  Fermi energy is set to be the reference energy on the horizontal axis.  (B-C) Corresponding Wannier function isosurfaces of the occupied orbitals (B) $f_{z^3}$ and (C) $f_{y(3x^2-y^2)}$, where blue and yellow represent positive and negative values, separately. The isosurface level is chosen at 20\% of the maximum absolute value. The $xyz$ coordinates represent the local coordinates used to define the orbitals of Pr. (D-F) DOS and Wannier functions of Pr $f$-orbitals on the Ba site. The local coordinates are rotated compared to (A-C). }
\label{fig:Fig5_f_orbs}
\end{figure}
\begin{figure*}[t]
\centering
\includegraphics[width=\linewidth]{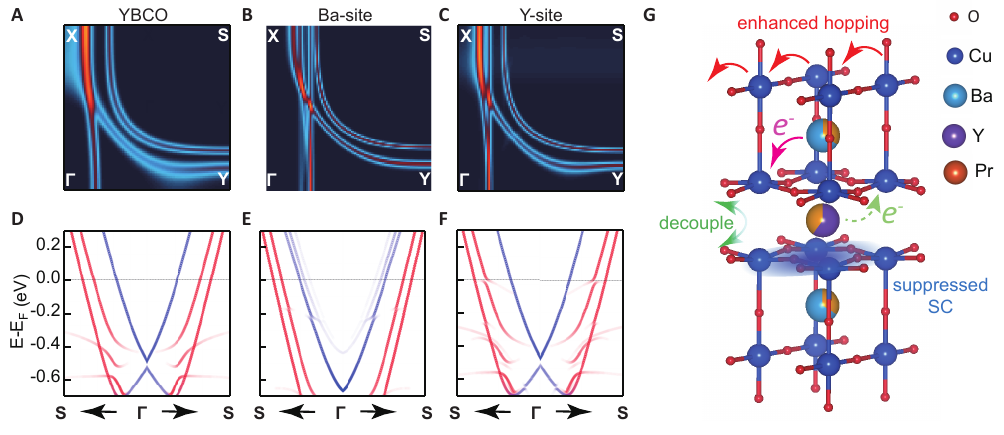}
\captionsetup{width=\linewidth}
\caption{\justifying Dual site occupancy effects of Pr substitution. (A-C) DFT Fermi surfaces for pristine, Ba-site 33\% and Y-site 33\% Pr substitution, respectively. (D-F) Nodal band structures. Blue (red) lines represent projections on chain Cu $d_{z^2}$ (planar Cu $d_{x^2-y^2}$) orbitals. (G) Schematic drawing of Pr substitution effects on two preferential sites.}
\label{fig:Fig6_DFT_FS}
\end{figure*}

The absence of FR states at $E_F$ compels a reassessment of Pr's role on other elements' valence states. One key structural effect with Pr-substitution is the flattening of the CuO$_5$ pyramid, which is generally considered to reduce copper valence with the lowered 3$d_{x^2-y^2}$ site energy~\cite{karppinen1992characterization,jin2024interlayer}. Another possible origin of additional electron doping is from partial Pr$^{3+}$ to Ba$^{2+}$ substitution, which was suggested in earlier studies but without conclusive experimental evidence~\cite{COLONESCU1999,PhysRevB.56.5512,PhysRevB.61.12404,10.1063/1.367709,PhysRevB.55.9160}. The presence of type 2 samples may relate to partial Ba-site substitution, which is evidenced by composition analysis, single crystal X-ray diffraction refinement (Table S2 and Fig.~\ref{supp-fig:XRD}), scanning tunneling microscopy (Fig.~\ref{supp-fig:STEM}), and disorder-induced chain band broadening effects (further discussed in DFT calculations). To understand the potential site-dependent Pr substitution effect, we model three systems: pristine YBa$_2$Cu$_3$O$_7$, Y$_{0.67}$Pr$_{0.33}$Ba$_2$Cu$_3$O$_7$ (Y-site substitution), and YBa$_{1.67}$Pr$_{0.33}$Cu$_3$O$_7$ (Ba-site substitution), using $3\times2\times1$ supercells with two Y/Ba atoms replaced by Pr. Structure relaxations reveal the lowest-energy configuration features Pr atoms aligned along Cu-O chains with antiferromagnetic (AFM) ordering. Ferromagnetic (FM) configurations cost $\sim$2 meV/Pr, which is consistent with the observed $\sim$17 K AFM transition in PrBa$_2$Cu$_3$O$_7$~\cite{PhysRevB.40.4453}. (Computational details in Methods and Text S2).

After full structural relaxation of both Y-site and Ba-site substituted systems, we find a striking result: the ground state of neither configuration exhibits occupied $4f_{z(x^2-y^2)}$ orbitals (Fig.~\ref{fig:Fig5_f_orbs}) -- the essential component for FR singlet formation. Instead, DFT+$U$ reveals that Pr $4f$ electrons predominantly occupy the $4f_{y(3x^2-y^2)}$ and $4f_{z^3}$ states, positioned approximately 4 eV below $E_F$. This orbital configuration aligns perfectly with crystal field expectations: as shown in Figs.~\ref{fig:Fig5_f_orbs}C and E, these specific $f$-orbitals minimize energy by orienting electron density away from neighboring oxygen atoms. In addition, Pr has a clear 3+ valence for both Y and Ba site replacement ($4f^2$ local configuration featuring a large energy gap).

To reconcile our findings with the LM model, we enforced occupation of the $4f_{z(x^2-y^2)}$ and $4f_{z^3}$ orbitals in Y$_{0.67}$Pr$_{0.33}$Ba$_2$Cu$_3$O$_7$ using occupation matrix control~\cite{allen2014occupation}. This constrained calculation converges to a meta-stable state $\sim$140 meV/Pr above the ground state energy, which indeed displays FR bands~\cite{PhysRevLett.70.3471} through antibonding Pr $4f_{z(x^2-y^2)}$-O $2p$ hybridization (Text S2). Extending this analysis to full Y-site substitution, we reproduced the LM-predicted band structure for PrBa$_2$Cu$_3$O$_7$~\cite{PhysRevLett.74.1000} (Text S2). Crucially, this configuration remains meta-stable, lying $\sim$328 meV/Pr above the true ground state. As with lower substitution concentrations, the authentic ground state features occupied $4f_{y(3x^2-y^2)}$ and $4f_{z^3}$ orbitals, rather than the low-energy $f$-states required for FR/LM hybridization. These results challenge the FR/LM mechanism as an explanation for superconductivity suppression in the low-doping regime (we discuss some other possibilities in our concluding remarks.)

Experimentally, Pr substitution nonetheless alters the low-energy electronic structure. Most notably, it leads to a strong decoupling of the CuO$_2$ bilayers and enhances hopping along the chains, as shown in Fig.~\ref{fig:Fig2_ARPES}. In Fig.~\ref{fig:Fig6_DFT_FS}, we present the DFT-calculated electronic structures of pure YBCO and Pr substituted YBCO at Ba and Y sites. Comparing YBCO (Fig.~\ref{fig:Fig6_DFT_FS}A, D), Ba-site Pr substituted YBCO (Fig.~\ref{fig:Fig6_DFT_FS}B, E), and Y-site Pr substituted YBCO (Fig.~\ref{fig:Fig6_DFT_FS}C, F), Pr substitution introduces three major changes in the electronic structure:
i) A reduction of bilayer splitting (more significant for Ba-site substitution, and to a lesser extent for Y-site substitution); ii) The emergence of a cascade of shadow chain bands in the Ba-site substituted sample; iii) Heavy electron doping, directly observed from the increased Fermi momentum of both chain and plane bands in the Ba-site substituted sample. We now discuss these changes and compare them with experimental results.

CuO$_2$ bilayer interaction is considered a key factor behind the enhanced $T_c$ in multilayer cuprates. The bilayer Josephson tunneling model has been proposed as a primary superconducting pairing mechanism in such systems~\cite{doi:10.1126/science.279.5354.1196,doi:10.1126/science.268.5214.1154}. Later, it was also suggested that the enhanced $T_c$ may arise from layer-dependent charge distribution, where layers with low and high carrier densities contribute strong pairing strength and phase stiffness respectively~\cite{kivelson2002making}. In both scenarios, bilayer coupling is essential for achieving high $T_c$. Given the relatively large bilayer coupling in YBCO, it serves as an excellent model system to investigate how bilayer splitting influences superconductivity. However, very few studies have demonstrated the ability to tune this coupling~\cite{fournier2010}.
Here, both our experimental results and DFT calculations show a reduction in bilayer splitting induced by Pr-substitution (Fig.~\ref{fig:Fig2_ARPES}L). The slower trend predicted from DFT indicates an additional correlation effect is likely at play, consistent with earlier reports in oxygenated YBCO~\cite{fournier2010}. Under this scenario, the disproportionate charge doping effect on the bonding vs antibonding bands (Fig.~\ref{fig:Fig2_ARPES}F) also predicts a robust superconducting pairing gap (due to nearly half-filled bonding band) with a weakening superfluid density (due to hole depletion on the antibonding band), which can be tested by further magnetic penetration depth measurements.

Now we turn to the loss of quasiparticle coherence in the chain band. First principles supercell calculations show that local structural distortions induced by Ba-site Pr substitution substantially reduce the lateral spacing between CuO chains by about 0.3~\AA, which in turn causes a $\sim 125$~meV shift in the onsite energy of the chain Cu $d{_z^2}$ Wannier orbital. This results in the appearance of shadow chain bands, as shown in Fig.~\ref{fig:Fig6_DFT_FS}B and E, which appear as smeared spectra experimentally (Fig.~\ref{fig:Fig2_ARPES}P and Q).
In contrast, Y-site Pr substitutions cause an order of magnitude smaller changes to the structure and onsite energy, leaving the chain bands largely unaffected, as seen in Fig.~\ref{fig:Fig6_DFT_FS}C and F. Meanwhile, the CuO$_2$ planes are less sensitive to Pr substitution due to their structural rigidity, as two additional oxygen atoms around each Cu atom reinforce the lattice, limiting its ability to distort. Pr substitutions cause atom displacements of up to 0.05~\AA~in the CuO$_2$ planes, leading to onsite energy changes of up to $\sim 25$~meV, again, an order of magnitude smaller than in the chains. These findings show that 1D CuO chains are far more susceptible (sensitive) to structural disruption (control) than their 2D counterparts. 

%Beyond these structural effects, strong correlation effects are also evident from the mismatch between DFT and experimental results. Comparing the DFT-calculated band structures (Fig.~\ref{fig:Fig5_DFT_FS}D for pristine and E for Ba-site doped), we see significant electron doping in both the chain and plane bands. However, experimental data (Fig.~\ref{fig:Fig2_ARPES}F) show heavy electron doping only in the plane bands. Furthermore, experimental results reveal significantly enhanced chain hopping, a feature not captured by DFT. The charge redistribution and band renormalization suggest that Pr doping modifies the chain–plane coupling through mechanisms beyond mere structural changes. This constitutes the first direct experimental evidence of chain–plane coupling modification in cuprates, establishing a promising testbed for future theoretical studies of 1D–2D crossover behavior.

To conclude, we present a comprehensive study of how Pr substitution modifies the electronic structure of YBCO. Fig.~\ref{fig:Fig6_DFT_FS}G summarizes the main findings of this work. First, superconductivity suppression in Pr substituted YBCO arises from substantial electron doping to the antibonding band. Our combined experimental and first-principles results also reveal the limitation of the FR/LM mechanism of hole localization effects, as no additional FR bands are seen near the Fermi level. Our calculations show that this model is a high-energy metastable configuration, whereas the predicted ground state for Pr replacement at the Y-site is solidly Pr$^{3+}$ with no predicted additional electron donation to the antibonding band. Looking to the future, we propose a few possibilities to explain the strong electron donation and $T_c$ suppression in Pr substituted YBCO:  (a) if the only effect of Pr substitution is to replace either Y or Ba and the DFT+U predictions for the ground-state electronic structure are correct, then the extent of electron donation must be linked with the fraction of Ba site replacement, which can be verified with high-resolution XRD; (b) incorporation of Pr has other structural effects (e.g., disorder, vacancies, interstitial ions) that act as electron donors; (c) there are novel many-body effects, missing from DFT+U theory, due to Pr replacement on the Y site that lead to a reduction of the effective hole count in the antibonding band (e.g., a many-body renormalization of spectral weight).
In addition to the above, we observe a rapid decoupling of the CuO$_2$ bilayers and enhanced electron hopping along the CuO chain, indicating significant modifications of both plane–plane and chain–plane coupling. The chain band shows extreme sensitivity to potential Ba-site disorder, while the planar bands remain robust. These results highlight Pr substituted YBCO as an ideal platform for investigating high-$T_c$ superconductivity and other correlated phenomena through site-specific electronic structure engineering of the CuO chains and CuO$_2$ planes.\footnote{All data for this work is available https://doi.org/10.6084/m9.figshare.30899429~.} 

\section*{Method}
\label{sec:method}
\subsection{Sample preparation}
Single crystals of nominal composition Pr$_x$Y$_{1-x}$Ba$_2$Cu$_3$O$_{7-\delta}$~\cite{MAPLE1992135} were synthesized following the procedure outlined in Ref.~\cite{PAULIUS1994255}. High-purity (99.99\%) Y$_2$O$_3$, Pr$_6$O$_{11}$, BaCO$_3$, and CuO powders were used as starting materials. Post-growth, the crystals were annealed in flowing oxygen to ensure full oxygenation and to optimize their superconducting properties. Atomic concentration is measured with Oxford-instrument EDS under JEOL 6610LV scanning electron microscopy and BRUKER XFlash 5060FQ Annular EDS detector under Hitachi SU8230 UHR CFE scanning electron microscopy. The superconducting transition temperatures ($T_c$) were characterized via magnetization measurements using a vibrating sample magnetometer integrated in a Quantum Design DynaCool Physical Property Measurement System.

\subsection{Single crystal X-ray diffraction}
Single crystal X-ray diffraction results are obtained through Rigaku XtaLAB Mini II system and Rigaku XtalLAB Synergy, Dualflex, Hypix single
crystal X-ray diffractometer at room temperature. Crystallographic
data acquisition was conducted employing $\omega$ scan methodology, utilizing Mo $K_\alpha$ radiation ($\lambda$ =
0.71073 $\AA$) emitted from a micro-focus sealed X-ray tube under operating conditions of 50 kV
and 1 mA. The determination of the experimental parameters, including the total number of runs
and images, was derived algorithmically from the strategy computations facilitated by the
CrysAlisPro software, version 1.171.42.101a (Rigaku OD, 2023). Subsequent data reduction
processes incorporated corrections for Lorentz and polarization effects. Integration of the
collected data, using the sphere model. An advanced numerical absorption correction was
implemented, leveraging Gaussian integration across a model of a multifaceted crystal~\cite{sheldrick2015}.
Moreover, an empirical absorption correction employing spherical harmonics was applied within
the SCALE3 ABSPACK scaling algorithm to refine the data further~\cite{sheldrick2015a}.

\subsection{Scanning tunneling electron microscopy}
Scanning tunneling electron microscopy imaging and EDS analysis were carried out with a Spectra Ultra microscope operated at 300 kV with a cold field emission gun. The EDS detector was Ultra-X EDS (silicon drift detectors with a collection solid angle of 4.45 srad). The probe semi-convergence angle was set at 30 mrad with a camera length of 110 mm and a probe current of 50 pA.

\subsection{None-resonance Inelastic X-ray Scattering}
The NIXS measurements were performed under ultrahigh vacuum at the High-Resolution Dynamics beamline P01 of PETRA-III at Deutsches Elektronen-Synchrotron (DESY, Germany). The incident x-ray beam energy was tuned using a Si(111) double-reflection crystal monochromator. The scattered photons were analyzed by a 3$\times$4 array of spherically bent 
Si(660) crystal analyzers fixed to an energy of 9690 eV. The energy loss spectra were measured by continuously sweeping the monochromator from 9690 eV (the elastic line) to higher energies, thus scanning the energy transferred in the inelastic scattering process. The experimental resolution, which is estimated by the full width at half maximum of the elastic line, was measured to be $\sim$1.4 eV. Fixing the scattering angle to $2\theta=155^{\circ}$ yields a momentum transfer vector $\bm{q}=\bm{k}_{in}-\bm{k}_{out}\approx9.6~\text{\AA}^{-1}$. All samples were polished to reduce surface defects. All NIXS spectra presented in this work were measured at 20 K and were normalized by the spectral weight of the Compton background.

\subsection{ARPES measurement}
Synchrotron ARPES measurements were performed at beamline 5 of the Stanford Synchrotron Radiation Lightsource. A hemispherical electron analyser (DA30, Scienta) was used. The k$_z$ dependence was taken with photon energy varying from 30 eV to 80 eV (Fig.~\ref{supp-fig:Kz_nodal}). All measurements were done using linear horizontal polarization. The Fermi surface map was done at 87 eV. The detector nonlinearity was calibrated and corrected. The chemical potential of the sample and the energy resolution of the system were determined by fitting the Fermi edge of polycrystalline gold. An energy-independent background was determined using intensity far above the chemical potential and subtracted from the data.

\subsection{First principles calculations}
All DFT calculations were based on the Vienna ab initio simulation package (VASP) with the projector-augmented wave method~\cite{kresse1999ultrasoft}.  A relatively high plane-wave cutoff energy of 500 eV is used, and a relatively dense $4\times 6\times 4$ $k$-grid is used for $3\times 2\times 1$ supercells.  All results come with full structural relaxation where energies and forces are converged to $10^{-6}$ eV and $10^{-3}$ eV/\AA, respectively.  The generalized-gradient-approximation (GGA) with the semilocal Perdew–Burke–Ernzerhof (PBE) functional~\cite{anisimov1991band, perdew1996generalized} is used in all calculations.  In addition, we add $U_{\text{Cu}}=4$ eV for the Cu $d$ manifold following previous theoretical works~\cite{jin2024first, deng2019higher}.  Varying $U_{\text{Cu}}$ between 0-9 eV shows little effect on the YBCO band structure for the paramagnetic state ~\cite{PhysRevB.79.064519}.  For the $f$-orbitals of Pr, prior theoretical works used $U_{\text{Pr}}=5$-$10$ eV ~\cite{tavana2009electronic, ghanbarian2006, ruiz2022, PhysRevLett.74.1000}.  We find that varying $U_{\text{Pr}}$ within this range always shows insulating Pr bands at least 1 eV away from the Fermi level, which does not qualitatively affect our results.  The results in the main text come with $U_{\text{Pr}}=8$ eV.  
Maximally localized Wannier functions ~\cite{marzari1997maximally} consisting of all Cu-d, O-p, and Pr-f orbitals were extracted from our DFT calculations using Wannier90~\cite{pizzi2020wannier90}.
To enable direct comparison with experimental ARPES measurements of the Fermi surface, we employ the standard band unfolding technique ~\cite{ku2010unfolding,brouet2012impact} for all electronic structures, which projects the band structure of a large supercell onto the Brillouin zone of the primitive unit cell. This approach has been shown to qualitatively reproduce spectral intensities observed in ARPES experiments across a wide range of materials~\cite{lin2011one, medeiros2014effects, tomic2014unfolding, zhu2018quasiparticle, jin2024first}.

\section{acknowledgement}We thank Bernhard Keimer for helpful discussions and the fully oxygenated YBCO samples of IXS measurements. J.Y., S.W., X.D., Z.K., and Y.H. acknowledge support from National Science Foundation under Grant DMR-2132343 and DMR-2239171. Z.J. and S.I. acknowledge support from National Science Foundation Grant No. DMR-2237469, National Science Foundation ACCESS supercomputing resources via allocation TG-MCA08X007, and computing resources from Yale Center for Research Computing. M.X. and W.X. acknowledge support from the the U.S. Department of Energy, Office of Science, Basic Energy Sciences under Contract DE-SC0023648. C.M., K.F., and M.B.M. acknowledge support from the U.S. Department of Energy, Office of Science, Basic Energy Sciences, under Grant No. DE FG02-04-ER46105. B.G. and A.F. acknowledge support from the National Science Foundation under Grant No. DMR-2145080. We acknowledge the Yale West Campus Materials Characterization Core. This research made use of the Chemical and Biophysical Instrumentation Center at Yale University. Materials characterization at Yale University is partially supported by the QuantumCT Quantum Regional Partnership Investments (QRPI) Award. We acknowledge DESY (Hamburg,
Germany), a member of the Helmholtz Association HGF, for the provision
of experimental facilities at PETRA III.
\bibliographystyle{apsrev4-2}
\bibliography{MAIN_v4.5_ref.bib}

\clearpage
\onecolumngrid
% ---- SI counters use S-prefix ----
\setcounter{figure}{0}
\setcounter{table}{0}
\setcounter{equation}{0}
\renewcommand{\thefigure}{S\arabic{figure}}
\renewcommand{\thetable}{S\arabic{table}}
\renewcommand{\theequation}{S\arabic{equation}}
% (Optional) S-prefixed sections:
% \setcounter{section}{0}
% \renewcommand{\thesection}{S\arabic{section}}

\begin{center}
  {\LARGE \textbf{Supplementary Information} \par}
  \vspace{0.7em}
  {\large Jinming Yang, Zheting Jin, Siqi Wang, Camilla Moir, Mingyu Xu, Brandon Gunn,\\
  Xian Du, Zhibo Kang, Keke Feng, Makoto Hashimoto, Donghui Lu, Jessica McChesney,\\
  Martin Sundermann, Hlynur Gretarsson, Shize Yang, Wei-Wei Xie, Alex Frano, Sohrab Ismail-Beigi, M.~Brian Maple, Yu He \par}
  \vspace{0.7em}
  {\today}
\end{center}
\vspace{1.0em}

% ---------------------------------------------------------------------------
% Text S1
% ---------------------------------------------------------------------------
\section*{Text S1. Tight Binding Model Fitting from ARPES}
\setcounter{NAT@ctr}{0}

The electronic spectra of multilayer cuprates have been extensively studied using angle-resolved photoemission spectroscopy (ARPES), where the effective interlayer couplings (EICs) between different CuO$_2$ planes give rise to band splitting~\cite{damascelli2003angle_sup}. Numerous studies have sought to quantify the strength of these EICs by fitting the observed band splitting into simplified tight-binding Hamiltonians, enabling the reproduction of experimental band dispersions based on these fits~\cite{chakravarty1993interlayer_sup, ideta2010enhanced_sup, ideta2021hybridization_sup, luo2023electronic_sup}. For simplicity, the most prominent fitting for cuprate superconductors is a Cu-only framework that employs local orbitals with $d_{x^2-y^2}$ ($d_{x^2}$) symmetry.
In prior literature, the EICs are usually described by an empirical formula
\begin{equation}
\Delta_{EIC} = t_0 + \frac{t_1}{4}\big(\cos(k_x a)-\cos(k_y b)\big)^2 ,
\label{equ:old}
\end{equation}
where $t_0$ and $t_1$ are two fitting parameters. However, this formulation fails to capture essential spectral features observed in some multilayer cuprate systems~\cite{luo2023electronic_sup}.  

\begin{figure}
\centering
\includegraphics[width=0.6\textwidth]{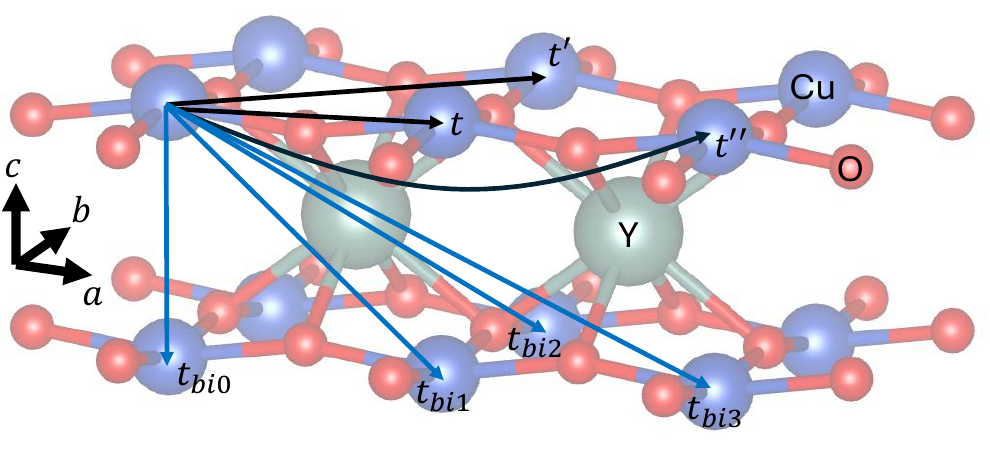}
\caption{Illustration of effective in-plane and interlayer hoppings in the tight-binding model of a YBCO CuO$_2$ bilayer.}
\label{fig:bilayer_hopping}
\end{figure}

In the main text, we successfully fit the ARPES spectrum with a more generic formula:
\begin{align*}
E_{\pm} = \epsilon_{\pm}-2t_{\pm} (\cos{k_x}+\cos{k_y})-4t_{\pm}^\prime\cos{k_x}\cos{k_y}-4t_{\pm}^{\prime\prime}(\cos{2k_x}+\cos{2k_y}) ,
\end{align*}
where $\pm$ represents antibonding (+) and bonding (-) bands, abbreviated as AB and BB. The band splitting $\Delta_{EIC} \equiv E_{+}-E_{-}$ quantifies the strength of the interlayer coupling between the CuO$_2$ planes. To extract a tight-binding model, the formula can be reformulated as:
\begin{equation}
\begin{aligned}
E_{\pm}& = \epsilon-2t (\cos{k_x}+\cos{k_y})-4t^\prime\cos{k_x}\cos{k_y}-4t^{\prime\prime}(\cos{2k_x}+\cos{2k_y}) \\
& \pm \Big[t_{bi0}-2t_{bi1} (\cos{k_x}+\cos{k_y}) -4t_{bi2}\cos{k_x}\cos{k_y}-4t_{bi3}(\cos{2k_x}+\cos{2k_y}) \Big],
\label{equ:TBmodel_energy}
\end{aligned}
\end{equation}
where $\epsilon \equiv (\epsilon_{+}+\epsilon_{-})/2$ and $t_{bi0} \equiv (\epsilon_{+}-\epsilon_{-})/2$. Other parameters ($t, t', t''$, etc.) are defined analogously. The energies of AB and BB are distinguished by the $\pm$ sign in the formula. 
From \eqref{equ:TBmodel_energy}, one can then read the following tight-binding model:
\begin{equation}
\begin{aligned}
\hat{H}& = \sum_{i l} \epsilon \,\hat{c}_{i l}^\dagger \hat{c}_{i l}
 -\sum_{\langle i,j\rangle l} t \,\hat{c}_{i l}^\dagger \hat{c}_{j l}
 -\sum_{\langle\!\langle i,j\rangle\!\rangle l} t^\prime \hat{c}_{i l}^\dagger \hat{c}_{j l}
 -\sum_{\langle\!\langle\!\langle i,j\rangle\!\rangle\!\rangle l} 2 t^{\prime\prime} \hat{c}_{i l}^\dagger \hat{c}_{j l} \\
&\quad +\sum_{i\langle l,l'\rangle} t_{bi0} \hat{c}_{i l}^\dagger \hat{c}_{i l'}
 -\sum_{\langle i,j\rangle \langle l,l'\rangle} t_{bi1} \hat{c}_{i l}^\dagger \hat{c}_{j l'}
 -\sum_{\langle\!\langle i,j\rangle\!\rangle \langle l,l'\rangle} t_{bi2} \hat{c}_{i l}^\dagger \hat{c}_{j l'}
 -\sum_{\langle\!\langle\!\langle i,j\rangle\!\rangle\!\rangle \langle l,l'\rangle} 2 t_{bi3} \hat{c}_{i l}^\dagger \hat{c}_{j l'} .
\label{equ:TBmodel_Hamiltonian}
\end{aligned}
\end{equation}
Here $i$, $j$ denote in-plane site indices, and $l$, $l'$ are layer indices. NN, NNN, and 3NN pairs are represented by $\langle\cdot,\cdot\rangle$, $\langle\!\langle\cdot,\cdot\rangle\!\rangle$, and $\langle\!\langle\!\langle\cdot,\cdot\rangle\!\rangle\!\rangle$, respectively.  
Figure~\ref{fig:bilayer_hopping} illustrates this generic tight-binding model on a YBCO bilayer of CuO$_2$ planes. Black arrows indicate in-plane hoppings; blue arrows indicate interlayer hoppings.  

Notably, the EICs in this model involve contributions from $t_{bi1}$, whose $k$-dependence is $\cos{k_x}+\cos{k_y}$, missing in the empirical formula \eqref{equ:old}. Recent studies have shown the importance of including this term in the fitting~\cite{photopoulos20193d_sup, jin2024interlayer_sup}. Microscopically, the effective $t_{bi1}$ hopping is mediated by interlayer hopping between O $p$ orbitals, which is large compared to $t_{bi2}$ and $t_{bi3}$ and cannot be ignored~\cite{jin2024interlayer_sup}. Consistently, the magnitude of $t_{bi1}$ fitted from experiment is larger than longer-ranged interlayer hoppings in the main text. 

% ---------------------------------------------------------------------------
% Text S2
% ---------------------------------------------------------------------------
\section*{Text S2. DFT Calculation Details}

\subsection*{Convergence of the Calculations}
All crystal and electronic structure calculations in this study were performed using VASP~\cite{kresse1996efficiency_sup, kresse1996efficient_sup}. Convergence is governed primarily by the plane-wave energy cutoff (ENCUT) and the $k$-point mesh density. We targeted convergence of energy differences within 1\,meV per dopant atom. All calculations used EDIFF=$10^{-6}$\,eV and EDIFFG=$-10^{-3}$\,eV/\AA. Gaussian smearing of 0.05\,eV was applied for SCF.

\begin{figure}
\centering
\includegraphics[width=\textwidth]{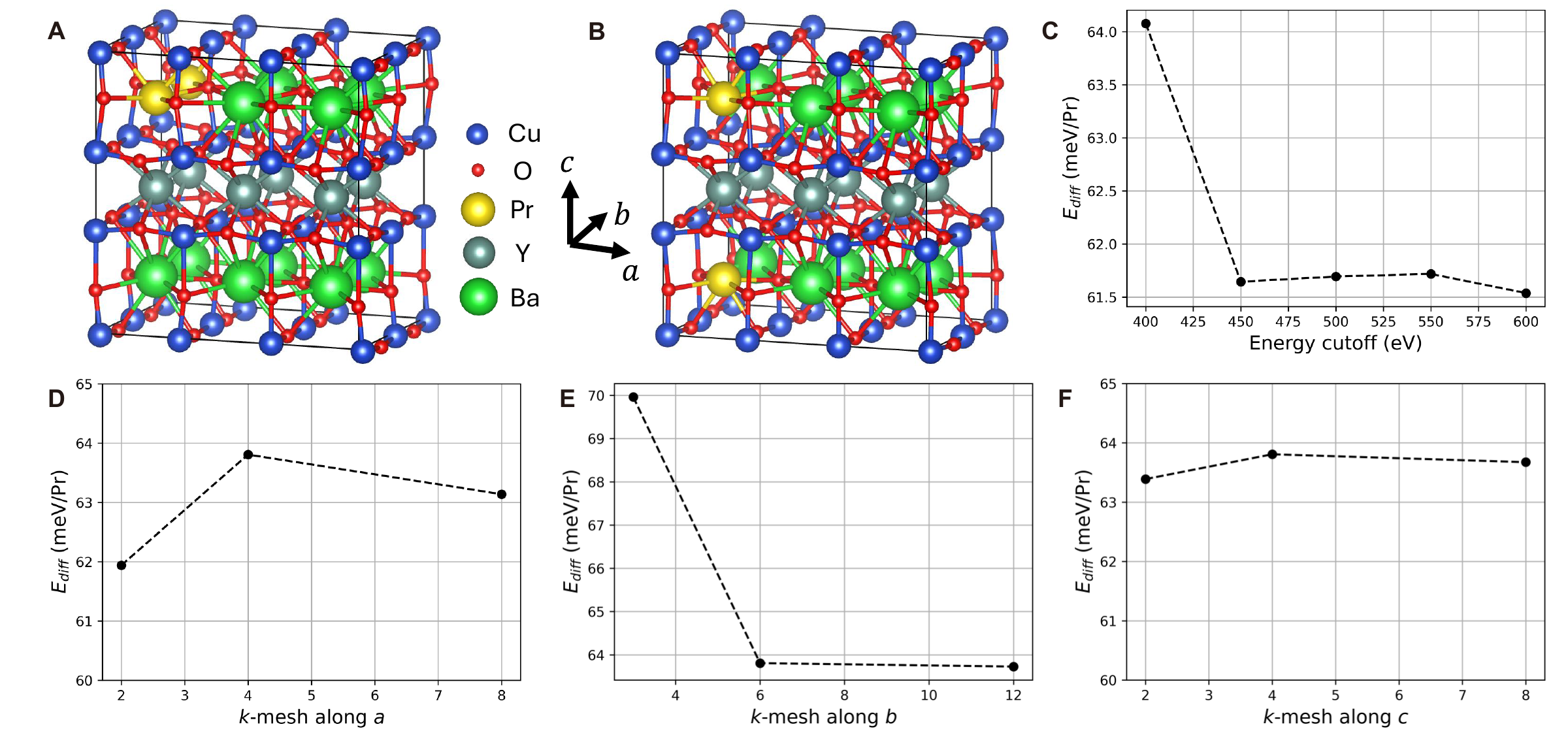}
\caption{Convergence of total-energy differences between dopant configurations vs.\ plane-wave cutoff and $k$-mesh density.
(A) Ba-site Pr dopants aligned along $b$.
(B) Ba-site Pr dopants aligned along $c$.
(C) $\Delta E$ per Pr between (A) and (B) vs.\ cutoff $E_c$.
(D–F) $\Delta E$ vs.\ $k$-mesh density along $a$–$c$. Default mesh is $4\times6\times4$ unless noted.}
\label{fig:convergence}
\end{figure}

Figure~\ref{fig:convergence}(a,b) show the lowest-energy and a metastable structure for distinct Ba-site Pr configurations. Figure~\ref{fig:convergence}(c–f) show convergence of their energy difference versus ENCUT and versus $k$-mesh density along three reciprocal directions. Based on these, we use ENCUT=500\,eV and a $4\times6\times4$ mesh throughout, sufficient to converge energy differences to within 1\,meV per Pr.

\subsection*{Magnetism and Effect of DFT+$U$ Corrections}
\begin{figure}
\centering
\includegraphics[width=0.65\textwidth]{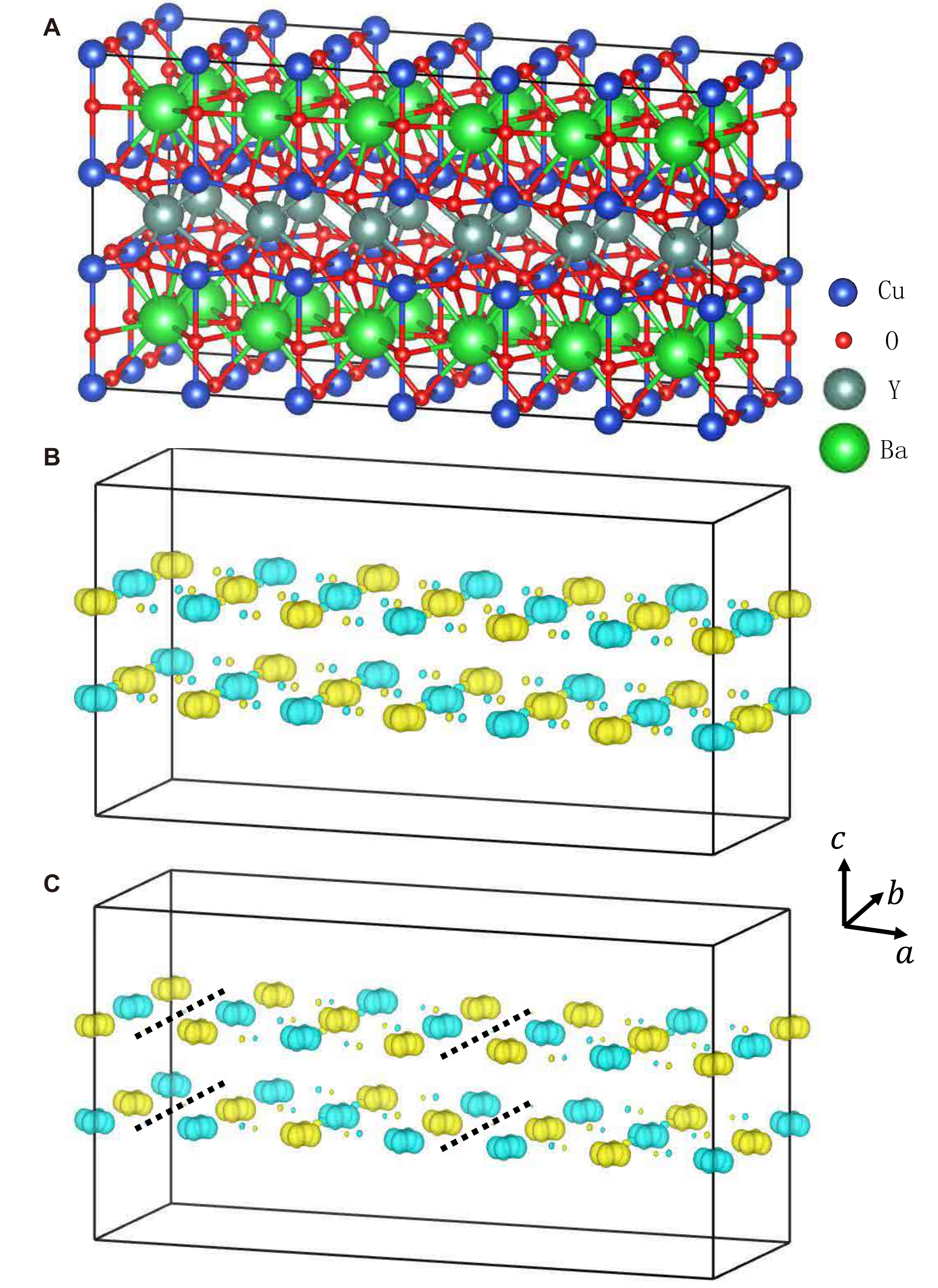}
\caption{(A) $6\times2\times1$ supercell. (B) Spin density isosurface of the G-AFM state. (C) Spin density isosurface of a stripe-ordered state; dashed lines mark domain walls.}
\label{fig:mag_order}
\end{figure}

Like many hole-doped cuprates, \YBCO{} exhibits competing magnetic orders on the CuO$_2$ planes~\cite{zhang2020competing_sup}; similar behavior occurs in Bi$_2$Sr$_2$CaCu$_2$O$_{8+x}$~\cite{PhysRevX.14.041053_sup}. Using PBE+$U$ with $U=4$\,eV, we identified many stripe-ordered states nearly degenerate with the G-AFM state. Figure~\ref{fig:mag_order} shows the spin densities for G-AFM and an example stripe state (“bond-centered” domain walls on O sites). The stripe state is only 1.2\,meV per planar Cu above G-AFM. Planar Cu moments are $\sim 0.41\,\mu_B$ (G-AFM), $\sim 0.36\,\mu_B$ near the domain wall, and $\sim 0.46\,\mu_B$ away from it—consistent with prior DFT work on YBCO7~\cite{zhang2020competing_sup}.

Because strong spin fluctuations are expected, a single ordered configuration is not an appropriate normal state. Prior work shows non-magnetic (NM) DFT bands agree well with ARPES Fermi surfaces~\cite{damascelli2003angle_sup, sobota2021angle_sup}, providing a reasonable paramagnetic proxy~\cite{PhysRevX.14.041053_sup}. Thus, NM Cu was used for band-structure comparisons in the main text.

\begin{figure}
\centering
\includegraphics[width=\textwidth]{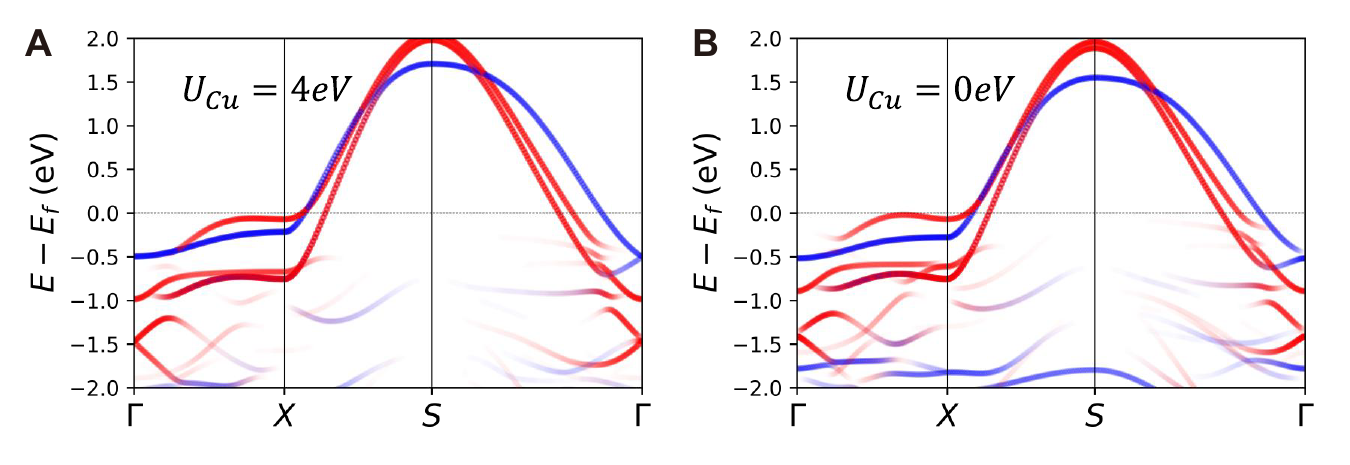}
\caption{(A) Projected bands of YBCO7 with DFT+$U$ ($U=4$ eV). (B) Projected bands without $U$. Red/blue indicate planar Cu $d_{x^2-y^2}$ / chain Cu $d_{z^2}$.}
\label{fig:diffU}
\end{figure}

We employ DFT+$U$ with $U=4$\,eV on Cu $d$ orbitals to reduce self-interaction errors~\cite{perdew1981self_sup}, following Refs.~\cite{yelpo2021electronic_sup, wang2006oxidation_sup, deng2019higher_sup, PhysRevX.14.041053_sup, jin2024interlayer_sup}. The NM band structure in YBCO7 is only weakly affected by moderate changes of $U$~\cite{jin2024interlayer_sup}, as illustrated in Fig.~\ref{fig:diffU}.

\subsection*{Stable or Metastable Crystals and Their Energies}
\begin{figure}
\centering
\includegraphics[width=\textwidth]{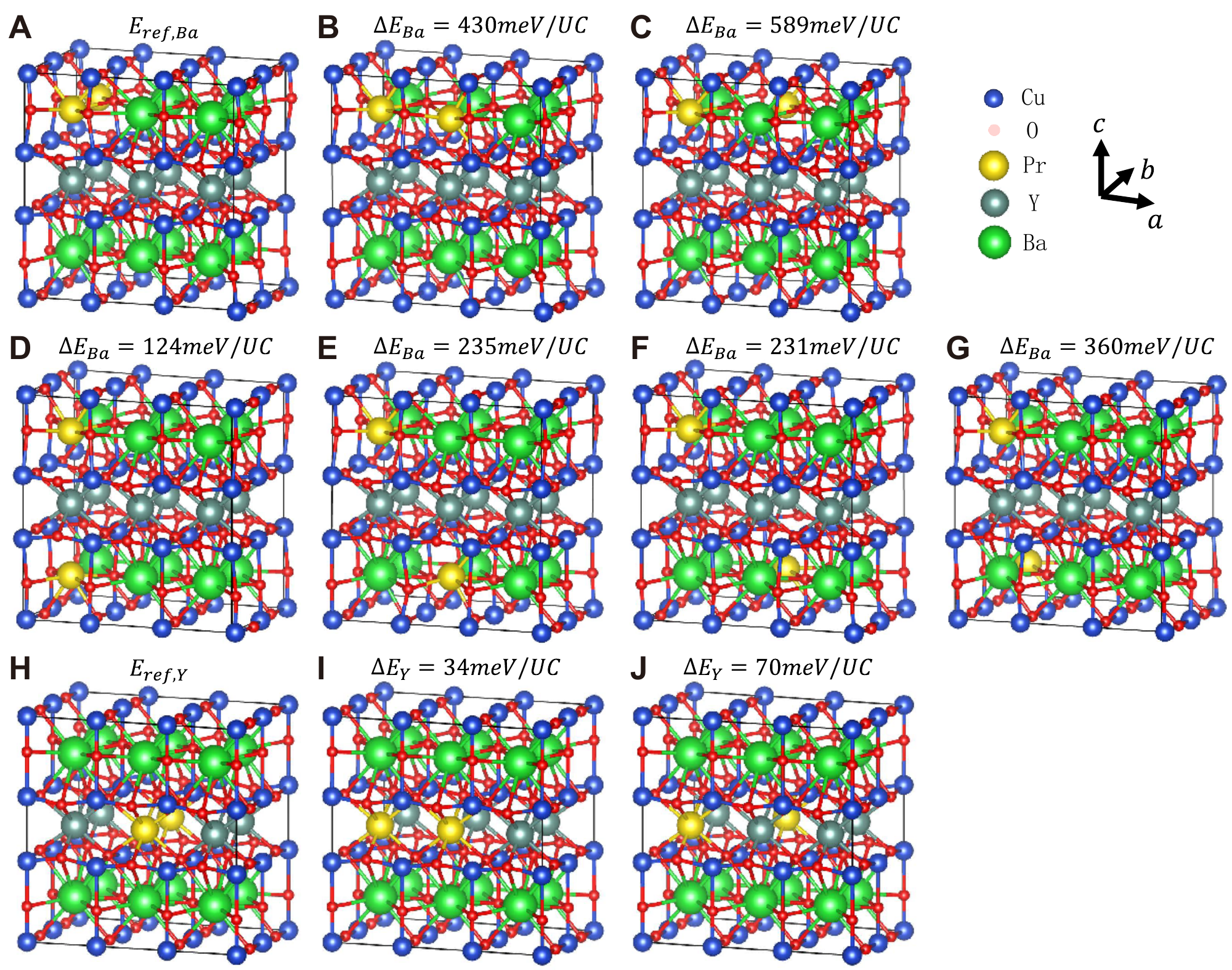}
\caption{Relaxed Pr dopant configurations in a $3\times2\times1$ supercell.
(A) Ba-site Pr aligned along $b$; lattice $(a,b,c)=(11.45,7.80,11.73)$\,\AA\ (lowest energy, reference).
(B–G) Other Ba-site metastable structures.
(H) Y-site Pr aligned along $b$; $(11.54,7.85,11.85)$\,\AA\ (lowest Y-site energy).
(I–J) Metastable Y-site structures.}
\label{fig:Pr_at_Ba}
\end{figure}

Metastable structures were frequently encountered during relaxation. We focus on Pr$_{0.33}$Y$_{0.67}$Ba$_2$Cu$_3$O$_7$ (Y-site) and Pr$_{0.33}$YBa$_{1.67}$Cu$_3$O$_7$ (Ba-site). In a $3\times2\times1$ supercell, two Pr atoms replace Y or Ba. Figure~\ref{fig:Pr_at_Ba} lists all relative positions investigated; all structures are fully relaxed. Planar Cu is set to NM; the two Pr spins are AFM-aligned. FM alignment is typically $\sim$1\,meV/Pr higher, consistent with $T_N\!\approx\!17$\,K~\cite{PhysRevB.40.4453_sup}.

\subsection*{$f$-orbitals}
\begin{figure}
\centering
\includegraphics[width=\textwidth]{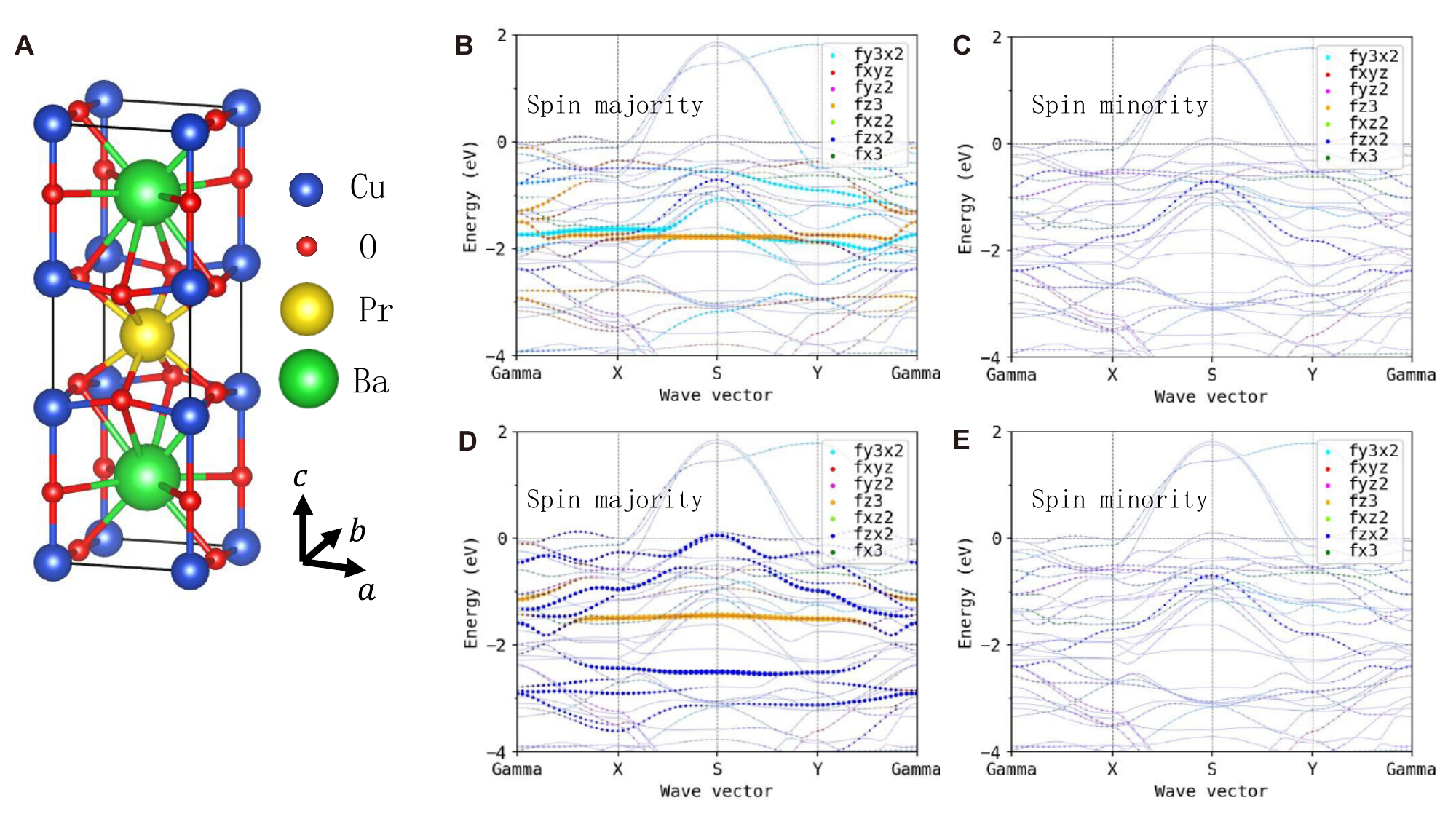}
\caption{(A) PrBa$_2$Cu$_3$O$_7$ primitive cell. (B–C) Ground-state spin-majority/minority: occupied $4f_{y(3x^2-y^2)}$ and $4f_{z^3}$. (D–E) Metastable FR/LM-like state with $4f_{z(x^2-y^2)}$ and $4f_{z^3}$; 328\,meV/Pr above ground state.}
\label{fig:f_orb}
\end{figure}

Following Liechtenstein–Mazin~\cite{PhysRevLett.74.1000_sup} for PrBa$_2$Cu$_3$O$_7$, we set $U_{\mathrm{Pr}}=6$\,eV and $U_{\mathrm{Cu}}=0$\,eV in the primitive cell (Fig.~\ref{fig:f_orb}a). Figures~\ref{fig:f_orb}(b,c) show the ground state with occupied $4f_{y(3x^2-y^2)}$ and $4f_{z^3}$ in spin-majority; other $f$ orbitals are empty and $\ge 1$\,eV from $E_F$. Using occupation-matrix control~\cite{allen2014occupation_sup} we explored 21 initial states; 18 distinct (meta)stable states were obtained (Table~\ref{tab:metastable}). The Fehrenbacher–Rice / Liechtenstein–Mazin configuration~\cite{PhysRevLett.70.3471_sup,PhysRevLett.74.1000_sup} is reproduced but is metastable, 328\,meV/Pr above the ground state (Figs.~\ref{fig:f_orb}d,e).

\subsection*{Wannierization}
\begin{figure}
\centering
\includegraphics[width=\textwidth]{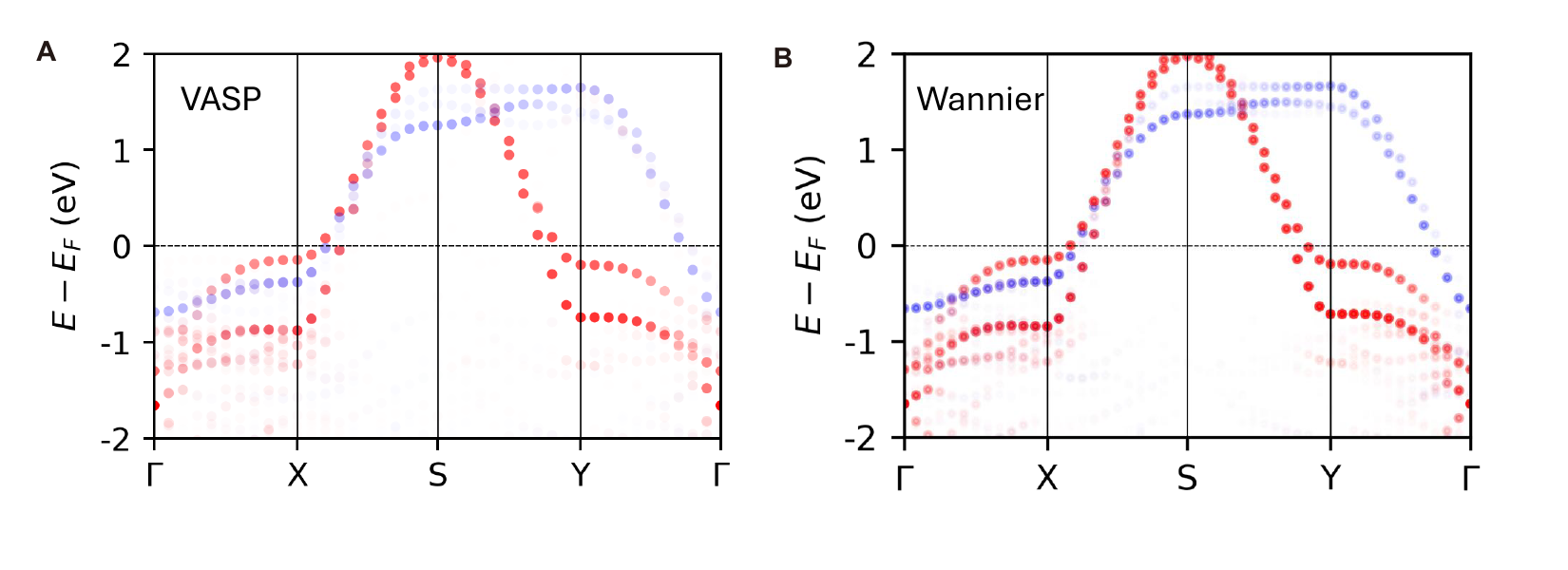}
\caption{Unfolded band structure of Ba-site Pr-doped YBCO7. (A) VASP. (B) Wannierized tight-binding. Opacity indicates orbital weight: red (planar Cu $d_{x^2-y^2}$), blue (chain Cu $d_{z^2}$), green (Pr $f$).}
\label{fig:wannier}
\end{figure}

Maximally localized Wannier functions~\cite{marzari1997maximally_sup} for Cu-$d$, O-$p$, and Pr-$f$ were constructed using \textsc{Wannier90}~\cite{pizzi2020wannier90_sup}. Figure~\ref{fig:wannier} compares the unfolded VASP bands with the tight-binding bands from Wannierization; excellent agreement confirms high-quality projections.

% ------------------------------- Tables ------------------------------------
\begin{table*}
\caption{Different stable/metastable states and their total energy relative to the ground state, $\Delta E$ in meV per Pr atom. Each state is identified by the two occupied spin-majority $4f$ orbitals.}
\begin{tabular}{c|cccccc}
\hline
$f$-orbitals & $f_{y(3x^2-y^2)}, f_{z^3}$ & $f_{z(x^2-y^2)}, f_{z^3}$ & $f_{xyz}, f_{z^3}$ & $f_{xyz}, f_{z(x^2-y^2)}$ & $f_{y(3x^2-y^2)}, f_{xyz}$ & $f_{y(3x^2-y^2)}, f_{z(x^2-y^2)}$ \\
$\Delta E$   & 0 & 328 & 196 & 482 & 189 & 523 \\
\hline
$f$-orbitals & $f_{yz^2}, f_{xyz}$ & $f_{z(x^2-y^2)}, f_{yz^2}$ & $f_{xz^2}, f_{z(x^2-y^2)}$ & $f_{xyz}, f_{z(x^2-y^2)}$ & $f_{x(x^2-3y^2)}, f_{z^3}$ & $f_{x(x^2-3y^2)}, f_{xyz}$ \\
$\Delta E$   & 491 & 158 & 330 & 489 & 1 & 192 \\
\hline
$f$-orbitals & $f_{x(x^2-3y^2)}, f_{z(x^2-y^2)}$ & $f_{xz^2}, f_{y(3x^2-y^2)}$ & $f_{yz^2}, f_{y(3x^2-y^2)}$ & $f_{yz^2}, f_{x(x^2-3y^2)}$ & $f_{yz^2}, f_{xz^2}$ & $f_{x(x^2-3y^2)}, f_{xz^2}$ \\
$\Delta E$   & 525 & 183 & 192 & 36 & 553 & 190 \\
\hline
\end{tabular}
\label{tab:metastable}
\end{table*}

\begin{table*}
\centering
\caption{Sample properties: doping level, $T_c$, and primary site occupancy.}
\begin{tabular}{c|ccccc}
\hline
Pr content (\%) & 0 & 5 & 15 & 12 & 28 \\
\hline
onset $T_c$ (K) & 91 & 91 & 84 & 63 & 53 \\
\hline
hole doping     & 0.31 & 0.34 & 0.22 & 0.18 & 0.11 \\
\hline
site occupancy (SCXRD) &  -- & Y site & Y site & Ba site & Ba site \\
\hline
site occupancy (EDX)   &  -- & Y site & Y site & Ba site & Y site (10\%) and Ba site (18\%) \\
\hline
\end{tabular}
\label{tab:samples}
\end{table*}

\begin{table*}
\centering
\caption{Tight-binding model fitting results (energies in meV).}
\begin{tabular}{c|cccc}
\hline
sample & $\epsilon_\pm$ & $t_\pm$ & $t'_\pm$ & $t''_\pm$ \\
\hline
Pr 0\% AB      & 387$\pm$2 & 436$\pm$2 & $-346\pm2$ & 24.4$\pm$0.2 \\
Pr 0\% BB      & 191$\pm$1 & 348$\pm$1 & $-333\pm2$ & 28.4$\pm$0.3 \\
\hline
Type 1 5\% AB  & 588$\pm$8 & 650$\pm$8 & $-531\pm8$ & 34.1$\pm$0.7 \\
Type 1 5\% BB  & 217$\pm$4 & 369$\pm$5 & $-353\pm7$ & 25$\pm$1 \\
\hline
Type 1 15\% AB & 244$\pm$5 & 307$\pm$7 & $-202\pm6$ & 25$\pm$1 \\
Type 1 15\% BB & 70$\pm$2  & 195$\pm$4 & $-156\pm5$ & 29$\pm$1 \\
\hline
Type 2 12\% AB & 224$\pm$7 & 287$\pm$7 & $-216\pm8$ & 6$\pm$1 \\
Type 2 12\% BB & 57$\pm$2  & 191$\pm$3 & $-120\pm4$ & 32$\pm$1 \\
\hline
\end{tabular}
\label{tab:samples_TBM}
\end{table*}

% ------------------------------- Figures -----------------------------------

\begin{figure}
\centering
\includegraphics[width=\textwidth]{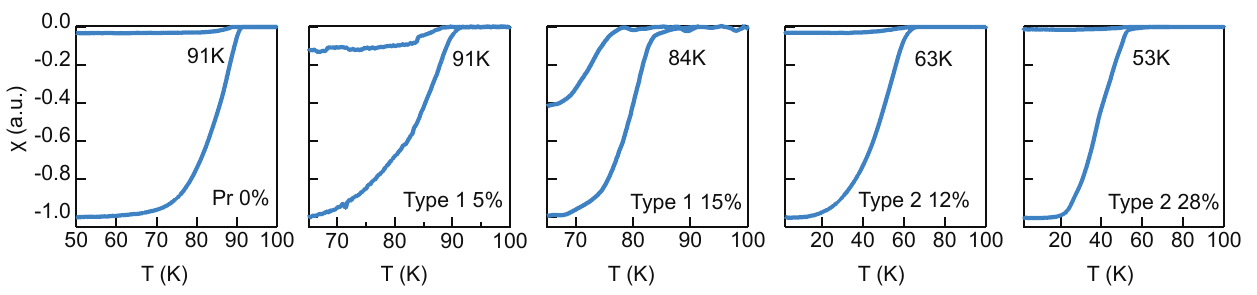}
\caption{Superconducting transition: magnetic moment under $H=50$\,Oe (out-of-plane).}
\label{fig:MPMS}
\end{figure}

\begin{figure}
\centering
\includegraphics[width=\textwidth]{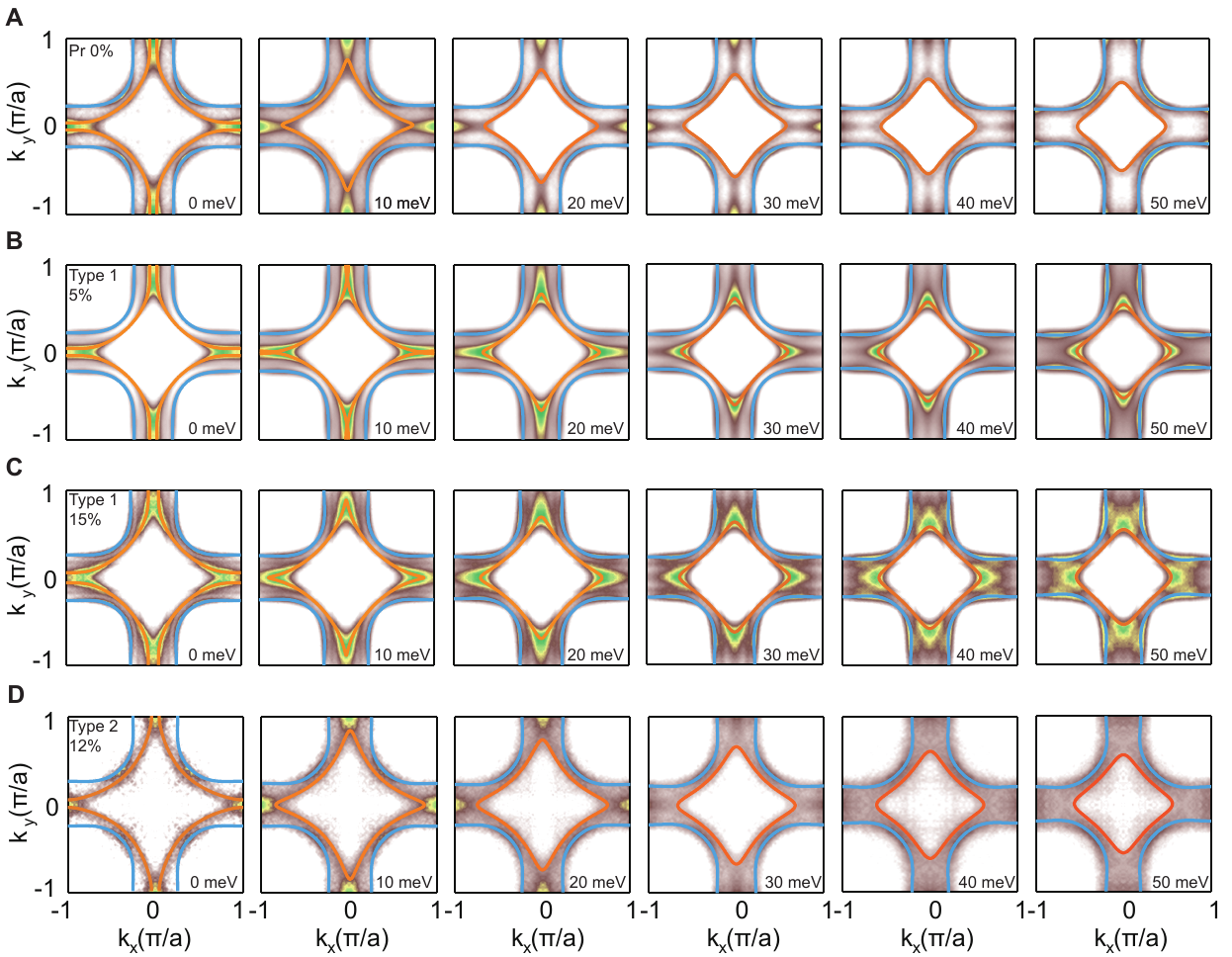}
\caption{Tight-binding fittings. Constant-energy maps for (A) Pr 0\%, (B) Type-1 5\%, (C) Type-1 15\%, (D) Type-2 12\% at 0–50\,meV. Orange: BB $k_F$; cyan: AB $k_F$.}
\label{fig:TBMfitting}
\end{figure}

\begin{figure}
\centering
\includegraphics[width=0.8\textwidth]{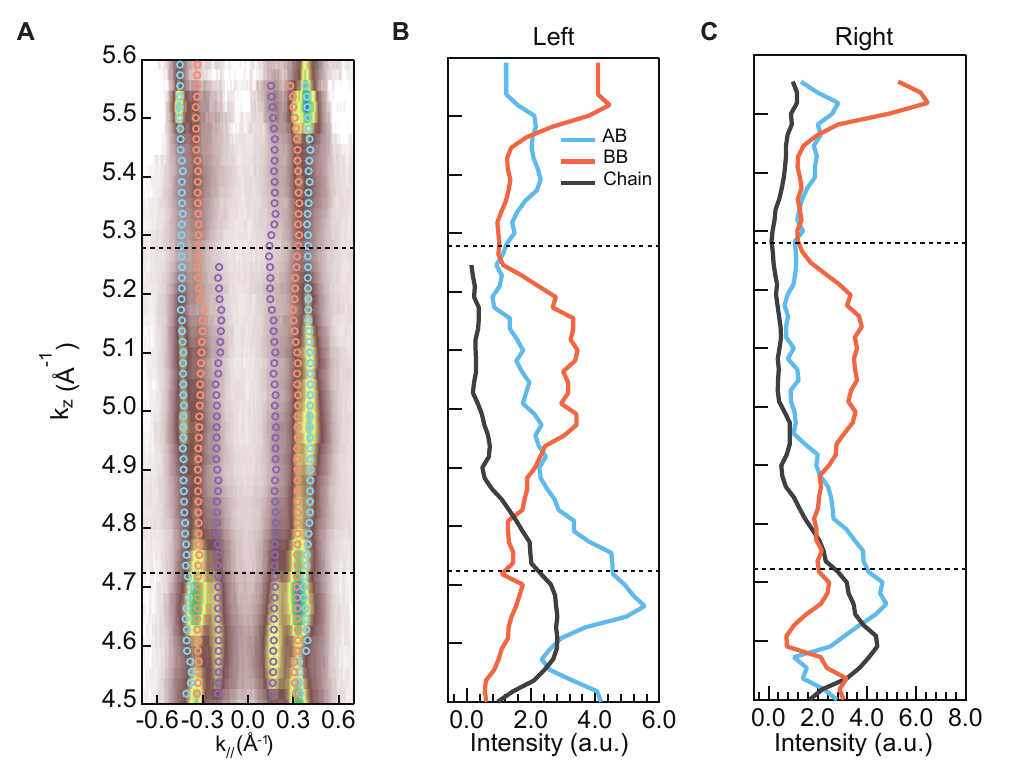}
\caption{$k_z$ dispersion along the nodal direction (30–80\,eV photons). (A) $E_F$ map vs.\ $k_\parallel$ and $k_z$; MDC peak positions for AB (cyan), BB (orange), chain (purple). (B–C) Band-specific $k_z$ intensity for left/right branches.}
\label{fig:Kz_nodal}
\end{figure}

\begin{figure}
\centering
\includegraphics[width=0.8\textwidth]{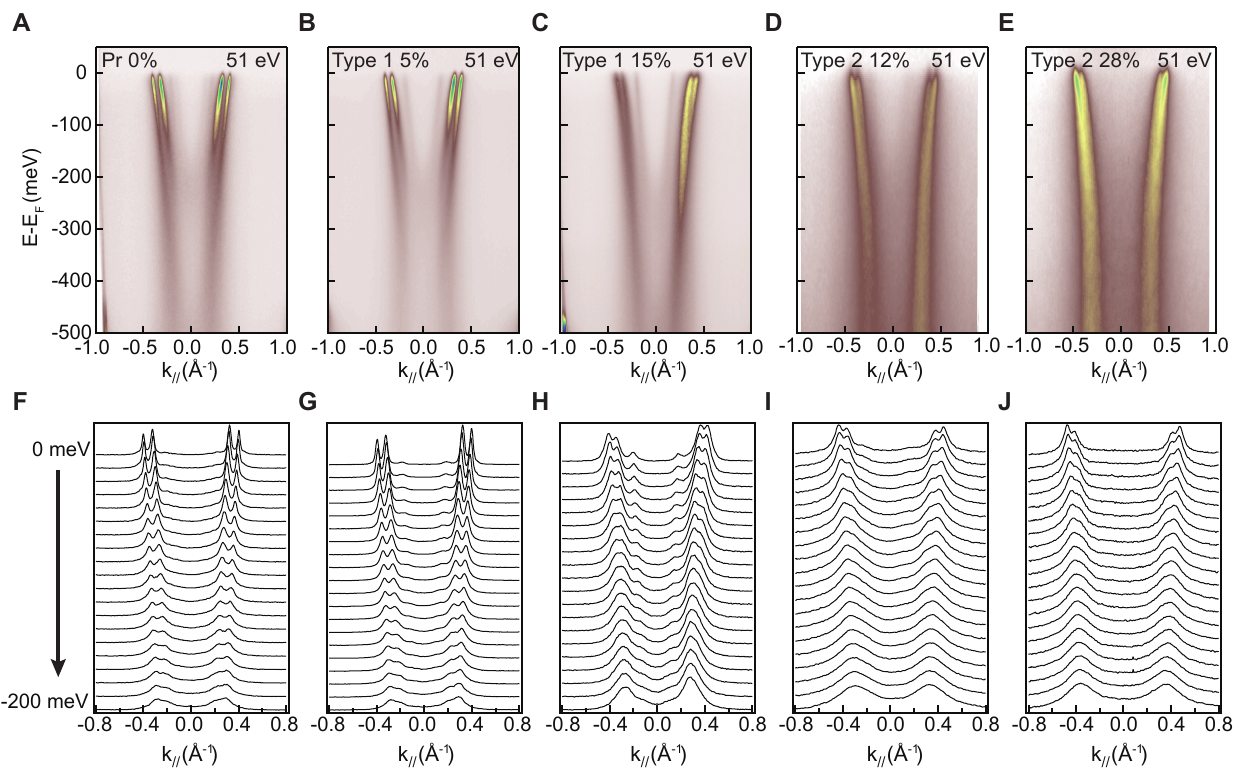}
\caption{(A–E) Nodal dispersions vs.\ photon energy highlighting plane bands for different Pr dopings. (F–J) Stacked MDCs from $E_F$ to 200\,meV.}
\label{fig:Bilayer_nodal}
\end{figure}

\begin{figure}
\centering
\includegraphics[width=0.8\textwidth]{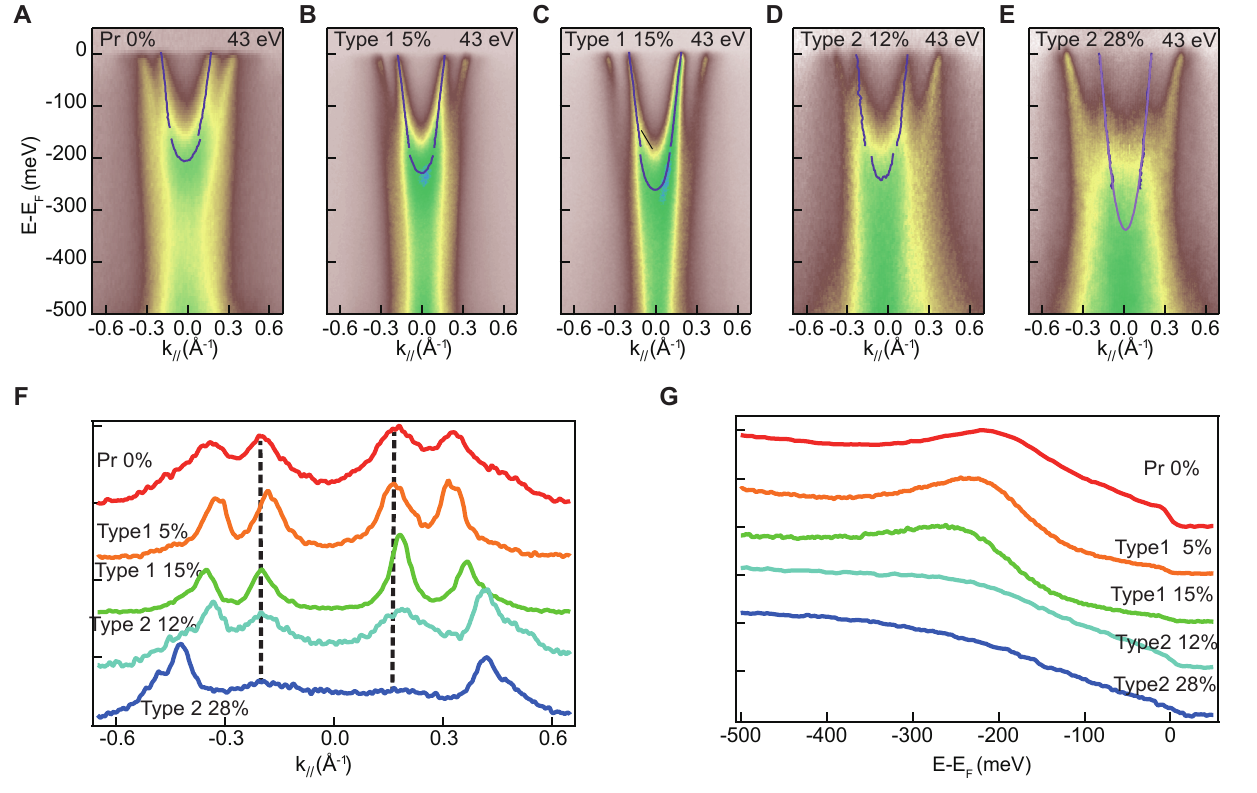}
\caption{(A–E) Nodal dispersions highlighting plane bands; fitted chain-band dispersion overlaid. (F) $E_F$ MDCs. (G) EDCs at $\Gamma$.}
\label{fig:Chain_nodal}
\end{figure}

\begin{figure}
\centering
\includegraphics[width=0.6\textwidth]{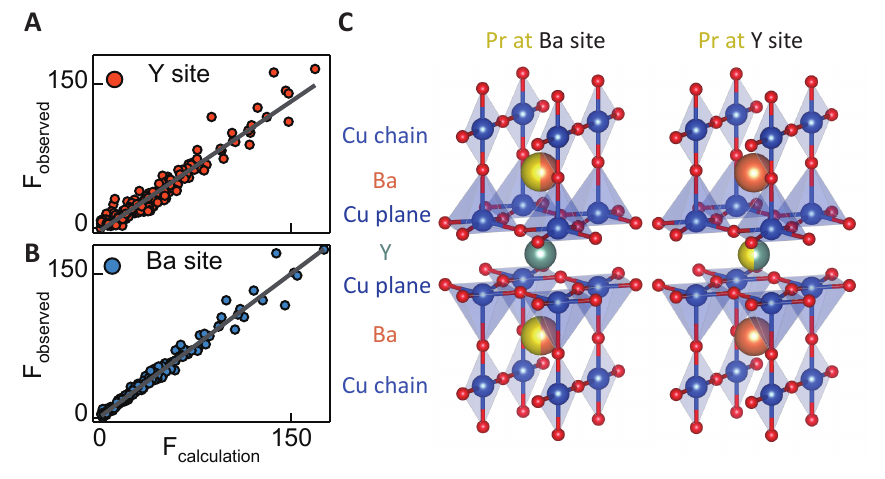}
\caption{Single-crystal XRD refinement of site occupancy. (A,B) Observed vs.\ refined structure factors for (A) Y-site and (B) Ba-site Pr substitution. (C,D) Structure models for (C) Y-site and (D) Ba-site substitution.}
\label{fig:XRD}
\end{figure}

\begin{figure}
\centering
\includegraphics[width=0.8\textwidth]{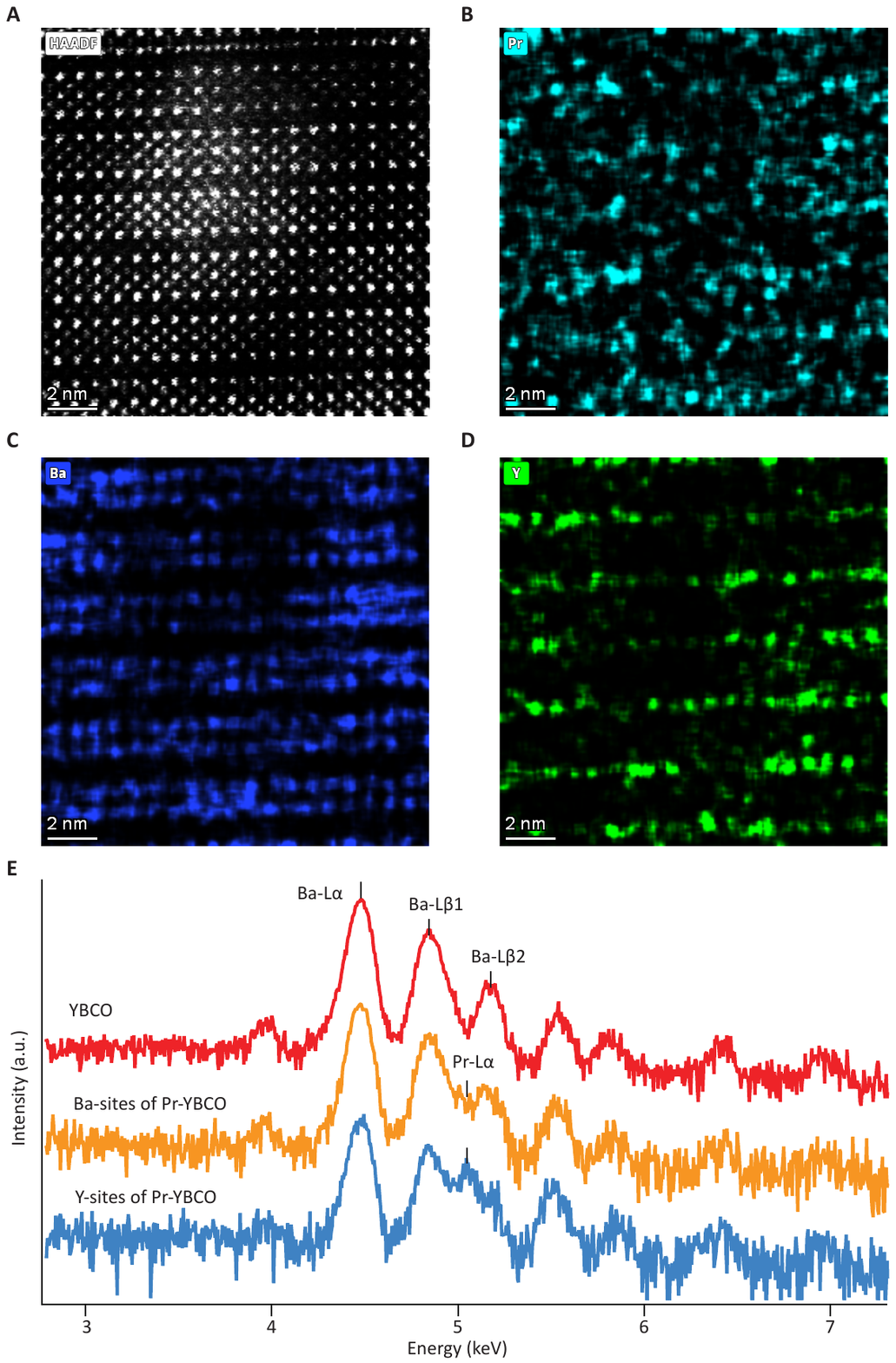}
\caption{HAADF-STEM and EDS for the Pr 12\% substituted YBCO: (A) HAADF-STEM image; (B) Pr; (C) Ba; (D) Y EDS maps. (E) EDS spectra for pristine YBCO (red) and Pr-YBCO (orange/blue). Pr-L$\alpha$ intensity appears at both sites, indicating Ba-site occupancy in addition to Y-site.}
\label{fig:STEM}
\end{figure}

% Ensure floats are processed before bibliography

% ---------------------------- Bibliography ---------------------------------
\clearpage
% Replace 'pnas-sample' with your .bib filename if different
\input{output_sup.bbl}

\end{document}